\documentclass[prb,url]{revtex4}
\usepackage{appendix}
\usepackage{graphicx,pdflscape,epsf,epsfig,amsmath}
\usepackage{alltt,dsfont}
\newcommand{\mats}[1]{{ i  \omega_{#1}}}
\newcommand{\Delslash}{{\bf \mathcal{D}}}
\newcommand{\llabel}[1]{\label{#1} }
\newcommand{\ups}{\Upsilon}
\newcommand{\iden}{ \mathds{ 1}}
\newcommand{\loc}{{\text{\tiny \mbox{loc}}}}
\newcommand{\lll}{\langle \langle}
\newcommand{\Del}[2]{\frac{\delta}{\delta \V_{#1}^{#2}}}
\newcommand{\rrr}{\rangle \rangle}
\newcommand{\A}{{\cal A}}

\newcommand{\jc}{\vec{J}}
\newcommand{\vecn}{\vec{\eta}}
\newcommand{\vecr}{\vec{r}}
\newcommand{\G}{{\cal{G}}}
\newcommand{\GL}{\hat{G}}
\newcommand{\GLO}{\hat{G}_0}
\newcommand{\GLI}{{\hat{G}}^{-1}}
\newcommand{\GLIO}{{\hat{G}_0}^{-1}}
\newcommand{\ttau}{{ \cal{T} }}
\newcommand{\tr}{{\text {tr} \ }}
\newcommand{\lind}{\chi_{\text{\tiny Lind}}}
\newcommand{\del}[1]{\delta_{#1}}
\newcommand{\htab}{\;\;\;\;\;}
\newcommand{\ua}{\uparrow}
\newcommand{\da}{\downarrow}
\newcommand{\X}[2]{X_{{#1}}^{#2}}
\newcommand{\si}{\sigma}
\newcommand{\sib}{\bar{\sigma}}
\newcommand{\ehat}{\hat{  E}}
\newcommand{\Jhat}{{\cal  J}}
\newcommand{\tJ}{\ $t$-$J$ \ }
\newcommand{\ess}{{\bf E}}

\newcommand{\self}{{{\cal{ S}}}}
\newcommand{\U}{{\cal U}}

\newcommand{\V}{{\cal V}}
\newcommand{\W}{{V}}

\newcommand{\beq}{\begin{equation}}
\newcommand{\eeq}{\end{equation}}
\newcommand{\barray}{\begin{eqnarray}}
\newcommand{\earray}{\end{eqnarray}}
\newcommand{\Din}{\Delta^{-1}}
\newcommand{\nn}{\nonumber}

\begin{document}
\title{  Extremely Correlated Quantum Liquids \footnote{Physical Review {\bf B 81},045121 (2010).}  }
\author{ B Sriram Shastry  }
\affiliation{Physics Department, University of California,  Santa Cruz, Ca 95064 }
\date{12 December 2009}
\begin{abstract}
Extreme correlations arise  as the limit of strong correlations,  when  the local interaction constant $U$ goes  to infinity.  This  singular limit transforms canonical  Fermions to  non canonical Hubbard type operators, with a specific graded Lie algebra replacing the standard anticommutators.
We are  forced  to deal with a fundamentally different and more complex lattice field theory.  We study  the   \tJ  model, embodying such extreme correlations. We 
formulate the picture of an extremely correlated electron liquid, generalizing  the standard Fermi liquid. This quantum liquid breaks no symmetries, and has specific signatures in various physical properties, such as the Fermi surface volume and the narrowing of electronic bands by spin and density correlation functions.
 We use  Schwinger's source field idea to generate equations for the Greens function for the Hubbard operators.  A local (matrix)  scale transformation 
in the time domain to a quasiparticle Greens function, is found to be optimal.  This transformation allows us to generate vertex functions that are guaranteed to 
reduce to the bare values for high frequencies, i.e. are ``asymptotically free''.  The quasiparticles are  fractionally charged objects,  and we find an exact
Schwinger Dyson equation for their Greens function, i.e. the  self energy is given explicitly in terms of the singlet and triplet particle hole vertex functions. We find a hierarchy of  equations for the vertex functions, and further  we obtain Ward identities so that systematic approximations are feasible.
 An expansion in terms of the density of holes measured from the Mott Hubbard insulating state follows from the
nature of the theory. A systematic presentation of the formalism is followed by some preliminary explicit calculations. We find a d-wave superconducting instability
at low $T$ that formally resembles  that found in the resonating valence bond (RVB) theory, but with a much reduced $T_c$.
\end{abstract}
\maketitle

\tableofcontents
\section{Introduction}
We present a theory of an extremely correlated quantum liquid (ECQL) in this report. We believe it to be  both necessary and useful to make a distinction 
between the class of systems studied here, and the more commonly addressed strongly correlated electron systems (SCES). The latter deals with models
such as the Hubbard model for transition metals or copper oxide systems, and Kondo or the periodic Anderson models as germane to heavy Fermi systems,
and also refers occasionally  to  the \tJ class of systems.
We use the term ECQL  exclusively for  systems such as the \tJ model\cite{early_tJ_1},  having   much stronger- even extreme  correlations built into them.
The origin of the difference is  the infinite Coulomb  repulsion at each site, where double occupancy is prohibited, not just discouraged as in the Hubbard model.
The result is that we must deal with a genuinely different problem, both physically and mathematically;  here the fundamental operators are no longer canonical 
Fermions but rather Hubbard operators. As we describe below, a systematic theory of the ECQL can be built, in parallel to the Fermi liquid (FL) theory for 
the more conventional interacting electron systems. The ECQL is described below as a coupled spin and charge liquid that breaks no symmetry, while it accommodates the extreme correlations of the \tJ class of models.  Its Greens functions and vertex functions can be defined in a  specific way that is different, 
and fundamentally more complex than in weak coupling (FL) models. In this formalism, the instabilities of the ECQL in charge, spin and superconducting channels 
can be  studied systematically. The extremely correlated quantum liquid described here, is a strong coupling entity and in contrast to the Fermi liquid, it cannot be related in an adiabatic fashion to  a free fermi  gas \cite{qhafm}.  Although the $t$-$J$ model is obtainable from a non interacting Fermi gas by turning on the $J$ parameter, one is also obliged to turn on the 
 Hubbard U term all the way to $U= \infty$.  Adiabaticity is lost in the passage to the infinite $U$ limit;   Appendix D illustrates  this breakdown of adiabaticity  within the context of the atomic limit of the Hubbard model, and outlines the general argument for the invalidity of the Luttinger Ward theorem 
 in the extreme correlation limit. The ECQL is expected to describe the Physics for a sufficiently large $U$, and not just $U=\infty$.

The theory of strongly correlated electronic systems, as opposed to the extremely correlated matter studied here, has received considerable attention in recent times.
A notable success has been the idea of large dimensionality where the dynamical mean field theory\cite{dmft-1,dmft-2} leads to a controlled set of calculations that 
are  useful as well as predictive.  The \tJ model Eq(\ref{hamiltonian}) has also  been studied\cite{lee}, but perhaps not so widely as the Hubbard model. The  origin
of the \tJ model is the  subject of several studies summarized in Ref\cite{tJ_model}. We emphasize that  it is not merely  a descendent of the  the Hubbard model 
(upon using  a large $U$ expansion), but rather 
has an independent origin via the down folding of multiband systems. Thus  $t$ and $J$ may be viewed as independent parameters, rather than  being fixed through $J= 4 t^2/U$ as in superexchange theory.  The possibility of an expansion of the physical  quantities in terms of  Mott Hubbard holes, 
i.e. the departure from  half filling, has
been anticipated in  1986 \cite{early_tJ_1}, but a systematic procedure remained undiscovered. Early studies  \cite{early_tJ_1,early_tJ_2}, have  struggled with the technical difficulties of the non canonical nature of the Fermions.
 The proposal of Anderson\cite{pwa} in 1987, that the High Tc systems are described by this model, has led to a revived 
and wider interest in this model.  Alternate techniques such as the
auxiliary field method have been employed  to deal with the constraints\cite{slave_operators,lee},  but  give divergent views of the outcome of extreme correlations. The present study is motivated by this  conundrum, as well as by the possibility of its applicability to certain systems found in nature. The predicted behaviour of the ECQL has certain specific signatures that  distinguish it from the FL, so that it should be relatively straightforward to verify its applicability to a given system.

The plan of the paper is as follows.  In Section \ref{sec2} we  present the calculation of the Greens function of the \tJ model, along with the background definitions of the Hubbard operators and the Schwinger framework. In Section \ref{sec-3}, the exact Schwinger Dyson equation is given after the sources are 
turned off. In Section \ref{app4}, we solve for the local Green's function.
 In Section \ref{vertices} we outline the computation of the vertices, with  detailed results given in Appendix \ref{vertices_all}.  In Section \ref{sec-6}
the Ward identities are developed and current vertices defined. In  Section \ref{sec-7}, we initiate the programme of successive approximations to the Greens function and list several consistent schemes. In Section \ref{sc} we present a calculation of the d-wave superconducting instability of the ECQL, using one of the consistent schemes for the Greens functions. In Section \ref{qp} we present a detailed picture of the quasiparticles which are argued to be fractionally charged. The argument for the changed Fermi surface volume is given further in Appendix \ref{hubbard-atomic}. We summarize the results in Section \ref{sec-10}. 

Appendix \ref{app2} and its subsections \ref{app-nu}, \ref{app3} and  Appendix \ref{app7} provide the details of the functional derivatives and the functions that arise in that process. Appendix \ref{app5} gives the details of the Nozi\`eres relations that play an important role in giving us rotation invariance of the theory. Appendix \ref{app8} provides the details of different  zero source limits that  are available. Appendix \ref{app9} contains details of  the connection between the quasiparticle
susceptibilities and the physical particle susceptibilities as well as some important sum rules that set the scales for the susceptibilities. Appendix \ref{vertices_all} gives all the vertices computed after throwing out higher order vertices, i.e. $\frac{\delta \Gamma}{\delta V} $. Finally Appendix \ref{app6} gives the conventions used for various Fourier transforms.

\section{  The bare and quasiparticle  Greens functions of the  \tJ model \llabel{sec2}}
We study the \tJ model given by
\beq
H= - \sum_{i,j} t_{i,j} \hat{c}^\dagger_{\vec{r}_j,\si} \hat{c}_{\vec{r}_i,\si} + \frac{1}{2} \ \sum_{i,j} \ J_{i,j} \{ \vec{S}_{\vec{r}_i} \cdot \vec{S}_{\vec{r}_j} - \frac{1}{4} {n}_{\vec{r}_i} {n}_{\vec{r}_j} +  {n}_{\vec{r}_i} \} - \mu \sum_i {n}_{\vec{r}_i}, \label{hamiltonian}
\eeq
where $\hat{c}^\dagger_{\vec{r}_i,\si}$ creates a  electron at the site $i$ which is {\em Gutzwiller projected}\cite{Gutz} to the subspace of single occupancy, and the other symbols have their usual meaning. The Gutzwiller projection creates the major technical challenge in this problem.  Projected electrons  no longer satisfy  canonical anti-commutation relations and  the  power of Wick's theorem\cite{wick} is lost, therefore a Dyson equation\cite{dyson} with
 straightforward expansion in terms of a free Greens function is not possible. Lost too are the appealing Feynman diagrams that encapsulate standard FL theory.
  The projected electrons satisfy instead,  a set of  graded Lie algebraic commutation relations.
 The latter are  compactly expressed in  the notation of Hubbard in terms of the $X$ operators that are summarized in Section \ref{app1}.  Section \ref{app1}  contains the details of the definitions of the Greens functions,  and of the equations of motion calculation, using the powerful technique invented by Schwinger, Martin   and their  school\cite{schwinger_school,kadanoff_baym,rajagopal}. We find  that despite  the non canonical nature of the Hubbard operators, we are able to obtain equations that are at the same level of  complexity as those of canonical Fermions,  with some unavoidable embellishments\cite{remark_1}.
 We therefore conclude that in the case of  projected electrons, the Schwinger-Dyson expansion\cite{rivers} is yet possible\cite{remark_2}.
This work studies the resulting equations and their consequences.

\subsection{ Basic framework using the Hubbard Operators \llabel{app1} }
Let us define the Hubbard operators acting upon each site as projected Fermi operators
\beq
\X{i}{\si_{1} \si_{2}}=|\si_{1}\rangle \langle\si_{2}|, \;\;\X{i}{0 \si_{1} } =| 0 \rangle \langle\si_{1}|,
\;\;\X{i}{\si_{1} 0}  =| \si_1 \rangle \langle 0 |,
\eeq
in terms of  the three possible states at any site  $|0 \rangle, |\uparrow \rangle, |\downarrow \rangle$.
The doubly occupied site is forbidden, and these operators do not connect  forbidden states with the allowed ones.
Thus an important statement of completeness at any site is the relation
\beq
\X{i}{00} = \iden - \sum_{\si} \X{i}{\si \si} . 
\eeq

The first member $  \X{i}{\si_{1} \si_{2}}$ is Bosonic while the other two are Fermionic  with respect to their 
commutation relations at different sites. The fundamental 
anticommutator for the Hubbard operators at {\em different sites } is most conveniently expressed as
\barray
\{ \X{i}{0 \si_{1}} ,\X{j}{\si_{2} 0} \} & = & \delta_{ij} [\delta_{\si_{1} \si_{2}} \X{i}{00} + \X{i}{ \si_{2} \si_{1}} ] \nn \\
& = & \delta_{ij} [\delta_{\si_1 \si_2} - \si_1 \si_2 \X{i}{\sib_1 \sib_2}  ].\llabel{hubbard-ops}
\earray

Also note the basic commutator
\beq
[ \X{i}{0 \si_3}, \X{j}{\si_{1} \si_{2}} ]= \del{ij} \del{\si_3 \si_{1}} \X{i}{0 \si_{2}}
\eeq
The Hubbard algebra at a given site is defined by
\beq
\X{i}{a b} \X{i}{c d} = \del{b c} \X{i}{ a d}
\eeq
In brief we may visualize these Hubbard operators in terms of familiar (canonical) electronic operators $c^\dagger_{\vec{r}_i,\si_a}$ creating an electron at site $i$ with spin $\si_a$, via  the non-linear constructs
\barray
\X{i}{\si_a, 0} = \hat{c}^\dagger_{\vec{r}_i,\si_a} & =& (1- n_{\vec{r}_i \ \sib_a}) \ {c}^\dagger_{\vec{r}_i,\si_a}  \nn \\
\X{i}{0 , \si_a}= \hat{c}_{\vec{r}_i,\si_a} & =& (1- n_{\vec{r}_i \  \sib_a}) \ {c}_{\vec{r}_i,\si_a} \nn \\
\X{i}{\si_a, \si_b}& = &  (1- n_{\vec{r}_i \ \sib_a}) \ {c}^\dagger_{\vec{r}_i,\si_a}\ {c}_{\vec{r}_i,\si_b}\   \llabel{canonical-electrons}
\earray
where $\sib_a = - \si_a $ here and throughout this paper. The factors of $(1- n_{\vec{r}_i \ \sib_a})$ get rid of doubly occupied sites. Although one can work with the $c$'s and project doubly occupied states out,   it is optimal to work with 
the $X$ operators. As Hubbard pointed out,    the manifold of lower Hubbard band states defined by the reduced Hilbert space excluding double occupation,  is mapped into itself
under  the action of $X$'s.

The Hamiltonian is expressed in terms of the $X$ operators by
\beq
H = - \sum_{i,j,\si} t_{ij} \X{i}{\sigma 0}\X{j}{ 0 \sigma} -\mu \sum_{i,\si}  \X{i}{\sigma \sigma} + \frac{1}{2} \sum_{i,j} J_{ij} \{ \vec{S}_i . \vec{S}_j - \frac{1}{4} n_i n_j \}  + \frac{1}{2} \sum_{i,j,\si}  J_{ij} \X{i}{\si \si}. \llabel{ham}
\eeq
The last (trivial) term is a shift of the chemical potential, and is added to make the equations more compact.
The  third ($J$ dependent) term  is rewritten as
\beq
V= \frac{1}{4} \sum_{ij} J_{ij} \{ P_{ij}- \X{i}{\si_1 \si_1} \X{j}{\si_2 \si_2} \} \htab = \frac{1}{4} \sum_{ij} J_{ij} \{ \X{i}{\si_1 \si_2} \X{j}{\si_2 \si_1} - \X{i}{\si_1 \si_1} \X{j}{\si_2 \si_2} \},
\eeq 
where $P$ permutes the spin indices.

\subsection{Calculation of the Greens function \llabel{greens}} 
The Greens function is defined as
\beq
\G_{\si_1 \si_2}[1,2]= -\langle T ( \X{1}{0 \si_1}(\tau_1) \X{2}{ \si_2 0}(\tau_2))\rangle,
\eeq
where $T$ is the time ordering symbol, and for an arbitrary operator $Q$ we define 
\beq
\langle Q \rangle = \frac{ Tr e^{- \beta H }  Q  }{Tr  e^{- \beta H }}.
\eeq
Thus, $\G$ is a  $2\times 2$ matrix in the spin space. The time dependence is given by
\beq
Q(\tau) \equiv e^{\tau H} Q e^{- \tau H}. \llabel{time_dependence}
\eeq

In further work we need to add a source term\cite{kadanoff_baym} via the operator $\A$  
\beq 
 \A = \sum_{j,\si_1,\si_2} \int_0^\beta \; d \tau \; \V_j^{\si_1 \si_2}( \tau ) \X{j}{\si_1 \si_2}(\tau), \llabel{action}
\eeq
with the {\em same } $\tau $ dependence of the operators as in  Eq(\ref{time_dependence}), and an arbitrary function of time $\V_j^{\si_1 \si_2}( \tau )  $ at every site. The Greens functions are now no longer time translation invariant, and are defined as 
\beq
\G_{\si_1 \si_2}[1,2]= -\frac{ Tr\left[  e^{- \beta H } T ( e^{-\A}\; \X{1}{0 \si_1}(\tau_1) \X{2}{ \si_2 0}(\tau_2))\right] }{Tr\left[  e^{- \beta H } T ( e^{-\A} )\right] }.
\eeq
This version of $G$ satisfies the Kubo Martin Schwinger boundary conditions\cite{kadanoff_baym} as may be readily verified
\beq
\G_{\si_1 \si_2}[r_1 0^-,r_2 0] = -\G_{\si_1 \si_2}[r_1 \beta,r_2 0], 
\eeq
so that one may perform a Fourier series with odd integer Matsubara frequencies only. We will use the convention that the Greens function is written as the difference of frequencies after setting the source $\A=0$, whereas for $\A \neq 0$ we will display a function of two separate times.  

More generally  for any variable we define a modified expectation 
\beq
\langle \langle Q(\tau_1,\tau_2,..) \rangle \rangle = \frac{ Tr\left[  e^{- \beta H } T ( e^{-\A}\; Q(\tau_1,\tau_2,..) \right] }{Tr\left[  e^{- \beta H } T ( e^{-\A} )\right] },
\eeq
with a compact notation that includes  the time ordering and the exponential factor automatically. Thus
\beq
\G_{\si_i \si_f}[i,f]= - \langle \langle X_i^{0 \si_i} \; X_f^{\si_f 0} \rangle \rangle.
\eeq
From this  the variation  of the  Greens function can be found from functional differentiation as
\beq
\frac{\delta}{\delta \V_j^{\si_1 \si_2}(\tau_1) } \lll Q(\tau_2) \rrr =  \lll Q(\tau_2) \rrr \; \lll \X{j}{\si_1 \si_2}(\tau_1)\rrr - \lll  \X{j}{\si_1 \si_2}(\tau_1) Q(\tau_2)  \rrr
\eeq
and we note the important commutator: 
\beq
[\X{i}{0 \si_{i}}, H] = - \sum_j t_{ij} \left[ \del{  \si_i \si_j }- { \si_{i} \si_{j}} \X{i}{\sib_{i} \sib_{j}}  \right] \X{j}{0 \si_{j}} - \mu \X{i}{0 \si_{i}} +
 \frac{1}{2} \sum_j J_{ij} \left[   \delta_{\si_i\si_j}- \si_i \si_j \X{j}{\sib_i \sib_j}  \right] \X{i}{0 \si_j}. 
\eeq
We note the similarity in form  between the $t$ and $J$ terms above, one can be transformed into the other by flipping the
spatial indices $i,j$ on the $X$ operators;   this symmetry  persists in the following  equations as well.
From this point  we will use an Einstein type convention,  we  sum over all  {\em internal repeated indices}, while leaving the external indices
fixed. By summing over an index, a spatial
 sum over the lattice and integration over imaginary time $0 \leq \tau_j \leq \beta$ is implied. The external indices (both space-time and spin) are recognizable since they appear in the LHS of all the equations. 

Let us compute the time derivative of the G. For this we need the derivative
\beq
\partial_{ \tau_i} T \left( e^{- \A} \X{i}{a,b}(\tau_i)\right)  = - T \left( e^{ - \A} [  \X{i}{a,b}(\tau_i), H ] \right) +   \V_i^{\si_1 \si_2}( \tau_i ) 
 T \left( e^{- \A} [ \X{i}{\si_1 \si_2}(\tau_i),\X{i}{a,b}(\tau_i)]\right). \llabel{eom_anyop}
\eeq
This follows from the definition of the time ordering and the form of $\A $. Using this we find: 
\barray
\partial_{\tau_i} \G_{\si_{i} \si_{f}}[i, f] & = & - \delta(\tau_i- \tau_f) \delta_{i,f} 
  \lll \left( \delta_{ \si_{i} \si_{f}}- { \si_{i} \si_{f}} \X{i}{\sib_{i} \sib_{f}} \right)   \rrr+
   \lll [  \X{i}{0 \si_{i}} (\tau_i),H] \; \X{f}{\si_{f} 0 }(\tau_f)   \rrr  \nn \\
&& -  \V_i^{\si_i \si_2}(\tau_1) \G_{\si_2 \si_{f}}[i, f].
\earray
We further  use the abbreviations, 
\begin{align}
\delta[i,j] & =  \delta_{i,j} \; \delta(\tau_i - \tau_j),
& t[i,j]  \  & =   t_{ij} \; \delta(\tau_i - \tau_j), \nn \\
J[i,j] &= J_{ij} \; \delta(\tau_i - \tau_j), &\V^{\si_{a} \si_{b}}_r &=  \V^{\si_{a}\si_{b}}_r[\tau_r]. \llabel{t-def}
\end{align}

We next introduce a  useful and convenient notion of    ``k-conjugation'' of any matrix $M$, denoted by $M^k$, such that $$(M^k)_{\si_1, \si_2}= \si_1 \si_2 M_{\sib_2 ,\sib_1}.$$ 
This is  the time reversal operation, but confined to the spin space indices and excludes transforming  the space-time indices.
The k-conjugation of a matrix in the spin space produces the transpose of its cofactor matrix, and thus the inverse of any matrix is proportional to its
conjugate, as in Eq(\ref{delta_def}). More explicitly we see that
\barray
M \cdot M^k &=& \det M \ \iden \nn \\
(M)^{-1} & = & \frac{1}{\det M} \ M^k, \llabel{k-conjugation}
\earray 
where $\iden$ is the identity matrix in the $2\times 2$ dimensional spin space.

We need an object $\Delta[i]$ that plays the role of a dynamical Gutzwiller factor in this theory. It is defined by
\barray
\Delta[i]&=& \iden -\G^k[i^-,i] \nn \\
\Delta_{\si_1 \si_2}[i]&=& \delta_{\si_1, \si_2} - \si_1 \si_2 \G_{\sib_2 \sib_1}[i^-,i]  \nn \\
\Delta^{-1}_{\si_1 \si_2}[i]&= & \frac{1}{\det \Delta[i]}\Delta^k_{\si_1 \si_2}[i]. \llabel{delta_def}
\earray
where the second and third lines follow from Eq(\ref{k-conjugation}). We also define a matrix functional derivative operator
\beq
D_{\si_1,\si_2}[r] = \si_1 \si_2 \Del{r}{\sib_1 \sib_2}.
\eeq
In terms of these, we find the equation of motion (EOM)
\barray
(\partial_{\tau_i} - \mu) \G_{\si_{i} \si_{f}}[i,f]  & = & - \delta[i,f]   \lll \delta_{\si_{i} ,\si_{f}} - { \si_{i} \si_{f}} \X{i}{\sib_{i} \sib_{f}}   \rrr  \nn \\
 &&  -  t[i,j]  \   \lll  \; \left ( \delta_{\si_{i} ,\si_{j}} - { \si_{i} \si_{j}} \X{i}{\sib_{i} \sib_{j}} \right )  \X{j}{0 \si_{j}}(\tau_i)  \; \X{f}{\si_{f} 0 }(\tau_f)   \rrr
 \nn \\
 &&   + \frac{1}{2}  J[i,j] \
    \lll  \; \left ( \delta_{\si_{i} ,\si_{j}} - { \si_{i} \si_{j}} \X{j}{\sib_{i} \sib_{j}} \right )  \X{i}{0 \si_{j}}(\tau_i)  \; \X{f}{\si_{f} 0 }(\tau_f)   \rrr
 \nn \\
&& -   \V_{i}^{\si_{i} \si_{j}}(\tau_i)\;\; \G_{\si_{j} \si_{f}}[i,f]. \llabel{eom_1}
\earray
We employ a  useful relation with an arbitrary operator  ${\cal Q}$:
\barray
 \lll  (\delta_{\si_a \si_b}- \si_a \si_b \X{a}{ \sib_a \sib_b } ) \ {\cal Q}  \rrr & = & (\Delta_{\si_a \si_b}[a] +
 D_{\si_a \si_b}[a] ) \ \lll  {\cal Q}  \rrr, \llabel{useful}
\earray
 to rewrite  Eq(\ref{eom_1})  in component form as
\barray
&& (\partial_{\tau_i} - \mu) \G_{\si_{i} \si_{f}}[i,f]   =  - \delta[i,f]   \Delta_{\si_i \si_f}[i] -   \V_{i}^{\si_{i} \si_{j}}(\tau_i)\;\; \G_{\si_{j} \si_{f}}[i,f] \nn \\
&& + \ t[i,j]    \left\{  \ (\Delta[i]+ D[i] ) 
\cdot \G[j,f] \ \right\}_{\si_i \si_f}   - \frac{1}{2}  J[i,j] \ \  \left\{  \ (\Delta[j]+D[j]) \cdot \G[i,f] \ \right\}_{\si_i \si_f}.  \nn \\ \llabel{eom_11}
\earray
This may  finally be written    compactly in matrix form as
\barray
(\partial_{\tau_i} - \mu) \G[i,f]  & = & - \delta[i,f]   \Delta[i] - \V_i \cdot \G[i,f]  - X[i,j] \cdot \G[j,f] -Y[i,j] \cdot \G[j,f] \llabel{eom_2},
\earray  
where we used the definitions
\barray
X[i,j]&=& - t[i,j]  \  D[i]+ \frac{1}{2} J[i,k] \  D[k] \delta[i,j] \nn \\
Y[i,j]&=& - t[i,j]  \  \Delta[i]+ \frac{1}{2} J[i,k] \  \Delta[k] \delta[i,j]. 
\earray
The space-time indices are displayed but the spin indices are  hidden in the above  matrix structure.

We next perform a scale (or local gauge) transformation  with a space time spin dependent factor described below. This scale transformation is a key step in our work,
and it is important to appreciate its motivation. If we work with the EOM Eq(\ref{eom_2}), the resulting vertex, i.e. schematically the object $-\frac{\delta}{\delta V} \G^{-1}$, turns out to have pathological  ``overhangs''. By this we mean that the vertex will  contain not only $\G$, but also   $ \G^{-1}$, i.e.  the putative  ``self energy'', and hence the resulting   Schwinger Dyson equation will be ill formed\footnote{ It is important to verify that 
  the self energy and vertices are well behaved at high frequencies $\omega$ (i.e. have the form $\sim c_1 + \ \frac{c_2}{\omega}$), and possess proper spectral representations.  When these conditions are not satisfied, as with equations formally following from those of $\G$ without the removal of the factor of $\Delta$ as in Eq(\ref{gauge_transform}),    we  denote the resulting equations as ill formed.}.  The origin of the difficulty is that the coefficient of $\delta[i,f]$ in the RHS of Eq(\ref{eom_2}) involves $\Delta[i]$ which is time dependent, and  essentially made up of $\G$. This in turn is a manifestation of the non canonical nature of the projected electrons.
The factor $\Delta[i]$ is physically understandable as arising from the spectral weight contained  in the lower Hubbard band, which is less than unity. On the other hand,    canonical Greens functions contain a spectral weight of unity. These observations are reflected in the coefficient of $\frac{1}{\omega}$ in the limit of high frequencies of the Greens functions changing from unity in the canonical case to $1-n_{\sib}$ for the \tJ model.

 We  resolve this difficulty  by a local space time dependent scale  transformation. With this, we   eliminate  this inconvenient factor of $\Delta[i]$ by a multiplicative process {\em in the time domain}. After the removal, we uncover  a new Greens function $\GL$ corresponding to   effective canonical electrons or quasiparticles (QP) Eq(\ref{canonical-electrons}) that lie underneath.
 One could  remove this factor $\Delta[i]$ in many equivalent ways, such as symmetrically or from the right, we choose a left sided transformation for maximal ease of computation\cite{symmetrize}.
Let us write
\beq
\G[i,f]= \Delta[i] \cdot \GL[i,f], \llabel{gauge_transform}
\eeq
so that the EOM Eq(\ref{eom_2}) becomes after some rearrangement:
\barray
 - \delta[i,f] \ \iden & = & \left\{  (\partial_{\tau_i} - \mu   +  \W_i  + \Phi_i ) \delta[i,j] \ \iden +  \Delta^{-1}[i] \cdot X[i,j] \cdot \Delta[j] 
   +  \Delta^{-1}[i] \cdot Y[i,j] \cdot \Delta[j]  \right\}.
  \GL[j,f] \nn \\
\W_i &=& \Delta^{-1}[i] \cdot \V_i \cdot \Delta[i] \nn \\
\Phi_i & = & \Delta^{-1}[i]\cdot ( \partial_{\tau_i}\Delta[i] ) 
 \llabel{eom_3}.
\earray
Here $\GL$ is the underlying canonical Greens function (with spectral weight unity), $\W$ is the transformed source field, and $\Phi$ arises from  the time derivative.
Appendix \ref{app7} summarizes its properties, and demonstrates that it is negligible 
on turning off the source terms.

\subsection{Calculation of the inverse Greens function}

We next turn to the task of finding the inverse Greens function. Using the methodology and results detailed in  Appendix \ref{app2}, (especially  Eq(\ref{mu_def}, \ref{mu_def_2})), and with $\vdots Q \vdots$ symbolizing a right  (i.e. normal) ordering of the functional derivative contained in the matrix operator $Q$,  we may rewrite Eq(\ref{eom_3}) as
\barray
 - \delta[i,f] \ \iden & = &   (\partial_{\tau_i} - \mu   +  \W_i  + \Phi_i ) \cdot \GL[i,f]  -t[i,j]  \  \Delta[j] \cdot \GL[j,f] + \frac{1}{2} J[i,k] \  
 \mu[i,k] \cdot \Delta[i] \cdot \GL[i,f] \nn \\
&&  -t[i,j]  \  ( \nu[i,j]+ \vdots\Delslash[i] \cdot \mu[i,j]\vdots )\ \cdot \GL[j,f]   \nn \\
&& + \frac{1}{2} J[i,k] \  (\mu[i,k] \cdot  \nu[k,i]+ \vdots\mu[i,k] \cdot \Delslash[k] \cdot \mu[k,i]\vdots )\cdot\GL[i,f] \llabel{eom_4}
  \earray

Here we denote
\barray
\mu[i,j] & = & \Delta^{-1}[i] \cdot \Delta[j] \ , \llabel{mu_def} \\
\nu[k,j]& = & \overline{\vdots \Delslash[j]\cdot\Din[j]\vdots\cdot\Delta[k] } \ , \llabel{nu-def} \\
\Delslash_{\si_1,\si_2}[i] & = & \si_1 \si_2 \frac{\delta}{\delta \W_i^{\sib_1 \sib_2}}. \llabel{identity-1-1} 
\earray
 The matrix products are  indicated by the center dots, and the terms under the overline symbol indicate the extent of terms over which 
the derivative acts.
 Using  the results detailed  in Appendix\ref{app2} Eq(\ref{identity_1}) 
and  Section\ref{app4} Eq(\ref{delta-sol-1}), we express various objects in terms of $\GL$ (rather than $\G$):
\barray
\Delta[j]& =& \frac{1}{\gamma[j]} \left( \iden - \GL^k[j^-,j] \right), \nn  \mbox{ where }\\
\gamma[j] & = &1-\det[\GL[j^-,j]]. \llabel{delta-sol-2}
\earray
  With these identifications, we have converted the problem to one only involving $\GL$ and $\W$  at this point, and jettisoned all reference to the original Greens function $\G$ and the original source $\V$.

Our next aim is to find an equation for the inverse  \footnote{This represents the dynamical inverse and  should not to be confused with a matrix inverse.} of $\GL$ defined through
\barray
\GL[i,j] \cdot \GLI[j,k] & = & \GLI[i,j] \cdot \GL[j,k] = \iden \delta[i,k] \nn \\
\GL_{\si_1,\si_2}[i,j] \GLI_{\si_2, \si_3}[j,k]&=& \delta_{\si_1,\si_3}\delta[i,k]. \llabel{ginv_def}
\earray
It is useful to define a ``susceptibility'' type three point object $\chi$ 
\barray
\chi^{\si_{a} \si_{b}}_{\si_{c} \si_{d}}[p,q;r] & \equiv & \frac{\delta \GL_{\si_{a}\si_{b}}[p,q]}{\delta \W_r^{\si_{c} \si_{d}}},  \llabel{chi}
\earray
and a vertex function $\Gamma$ 
\barray
\Gamma^{\si_{a} \si_{b}}_{\si_{c} \si_{d}}[p,q;r] & =& 
 -  \frac{\delta \GLI_{\si_{a} \si_{b}}[p,q]}{\delta \W_r^{\si_{c} \si_{d}}}, \;\;\;\mbox{so that}  \nn \\
\chi^{\si_{a} \si_{b}}_{\si_{c} \si_{d}}[p,q;r] & = & \GL_{\si_a \si_1}[p,x]\Gamma_{\si_c \si_d}^{\si_1 \si_2}[x,y;r]\GL_{\si_2 \si_b}[y,q] \llabel{chi_vertex}.
 \earray
The relations of these susceptibilities with the physical ones are detailed in Appendix \ref{app9}.
Let us right multiply Eq(\ref{eom_4}) by $\GLI[f,m]$ ( and sum over $f$)  so that
\barray
 && - \GLI[i,m]  =    (\partial_{\tau_i} - \mu   +  \W_i  + \Phi_i ) \ \delta[i,m] \ \iden 
 -t[i,m]  \  (  \nu[i,m] + \Delta[m])  \nn \\
&&  + \frac{1}{2} J[i,k] \  ( \mu[i,k] \cdot \nu[k,i] + \mu[i,k] \cdot \Delta[i] ) \ \delta[i,m] \nn \\
 && -t[i,j]  \    \overline{ \vdots\Delslash[i] \cdot \mu[i,j]\vdots \ \cdot \GL[j,f]} \cdot \GLI[f,m] 
   + \frac{1}{2} J[i,k] \  \overline{ \vdots\mu[i,k] \cdot \Delslash[k] \cdot \mu[k,i]\vdots \ \cdot \GL[i,f]} \cdot \GLI[f,m]  \nn \\
 \llabel{eom_5}
\earray
 More explicitly, the derivative excludes operating upon the factor $\GLI[f,m]$, since the overline excludes that term. 
The detailed calculation is  presented in Appendix \ref{app2}. 
We use  Eq(\ref{theta_def_2},\ref{theta_def}) and Eq(\ref{nu_def}) to rewrite this expression. We   split the Greens function into  the following form
with  $\GL_0$ as the bare quasiparticle Greens function and $\self$ as the quasiparticle self energy,
\barray
  \GL^{-1}[i,j] & = &  \GLIO[i,j] - \self[i,j],   \;\;\text{where}   \nn \\
 - \GLIO[i,j]  & \equiv &   (\partial_{\tau_i} - \mu   +  \W_i   + \frac{1}{2} (\sum_k J_{i,k} ) (1-\frac{n}{2})  )  \delta[i,j]       -t[i,j]  (1-\frac{n}{2}).
 \llabel{gsplit}
\earray
 The self energy is given by
\barray
 && \self[i,j]  =     \Phi_i   \delta[i,j] -t[i,j]  \  \{ (\Delta[j]  -(1- \frac{n}{2})) +  \nu[i,j] \}   \nn \\
&&  - t[i,k]  \   \Theta[i,k,j]   +  \ \delta[i,j] \frac{1}{2} J[i,k] \  \{ \mu[i,k] \cdot \Delta[i] -(1-\frac{n}{2})\} \nn \\
&& + \ \delta[i,j] \ \frac{1}{2} J[i,k] \   \mu[i,k]\cdot \nu[k,i] +  \frac{1}{2} J[i,k] \ \mu[i,k] \cdot \Theta[k,i,j] . \nn \\
 \llabel{schwinger_dyson_self_energy}
\earray
Regarding   nomenclature, we call the expression $\GL_0$ as the bare- rather than the unperturbed- Greens function, and the rest as self energy,  since there is no perturbation theory involved here.
The result Eq(\ref{gsplit},\ref{schwinger_dyson_self_energy}) is the exact Schwinger  Dyson  equation including the source field, and is central to the calculation of the vertices. Calculating the vertices proceeds by taking functional derivatives of Eq(\ref{schwinger_dyson_self_energy}), and leads to a standard  hierarchy of equations
for higher order vertices \cite{schwinger_school,rajagopal}.  We should also clarify that in contrast to the nomenclature of Fermi liquids,
 the quasiparticles as defined here are not infinitely sharp in energy,
they decay through relaxation processes that we describe via the imaginary part of their (analytically continued) self energy. Thus the quasiparticles of the ECQL have a coherent part as well as a 
incoherent background part that emerges in complete analogy to the electron Greens function in the FL.

\section{   Schwinger Dyson Equation and the Limit of vanishing sources \llabel{sec-3}}

Let us gather various objects here when the sources are turned off i.e. $V\to 0$.  We recover all the symmetries in this limit, 
translation invariance allows us to perform Fourier transforms in space and time, and rotation invariance leads to important simplifications of the vertices and susceptibilities. Due to the rotation invariance of the extremely correlated quantum liquid, the 
three triplet states of the particle hole pair are degenerate.  This leads to important  identities that parallel the relations in the Fermi liquid.  We denote these as  Nozi\`eres relations and   discuss them further in Appendix \ref{app5}, where the abbreviated superscripts and subscripts of the vertices are defined.
 Therefore as $V \to 0$,
\begin{align}
\G_{\si_1,\si_2}[i,j]& \rightarrow  \delta_{\si_1,\si_2} \G[i-j],
&\GL_{\si_1,\si_2}[i,j] & \rightarrow  \delta_{\si_1,\si_2} \GL[i-j], \nn \\
\Delta_{\si_1,\si_2}[k]& \rightarrow   \delta_{\si_1,\si_2}(1- \frac{n}{2}), 
&\mu_{\si_1,\si_2}[k,l]& \rightarrow  \delta_{\si_1,\si_2} 
\end{align}
and for any object $Q= \chi, \;\ups, \;  \mbox{or} \; \Gamma$ we have a spin decomposition:
\barray
Q_{\si_3,\si_4}^{\si_1,\si_2} &\rightarrow & \delta_{\si_1,\si_2} \delta_{\si_3,\si_4} \ \{\delta_{\si_1,\si_3} Q^{(1)} + \delta_{\si_1,\sib_3} Q^{(2)} \} + \delta_{\si_1,\si_3} \delta_{\si_2,\si_4} \delta_{\si_1,\sib_2} Q^{(3)} \nn \\
Q_t&=& Q^{(1)} -  Q^{(2)} = Q^{(3)}, \nn \\
Q_s&=& Q^{(1)}+  Q^{(2)}. 
\earray
The object $Q_s$ refers to the particle hole singlet channel, corresponding to a charge density variable and $Q_t$ to the particle hole triplet
channel, i.e. a spin density variable.

Let us note that in the limit of vanishing sources, the  physical  projected electron Greens function $\G$ and the quasiparticle Greens function
$\GL$ are simply related as
\barray
\G[i-j] & = & (1- \frac{n}{2}) \GL[i-j] \;\;\;\;\text{and} \nn \\
\G[\vec{k}, i \omega_n] & = & (1- \frac{n}{2}) \GL[\vec{k},i \omega_n].  \llabel{physical-greens}
\earray

The object  $\nu$   is now a  function of the difference $a-b$, and due to  the multiplying  factors $t[a,b], \; J[a,b]$ in the expression for $\GL$, we usually  need this at equal times $\tau_a=\tau_b$.
 We Fourier transform using the convention in Appendix \ref{app6} as 
\barray
{\nu}_{\si_1,\si_2}[a,b] &=&  \delta_{\si_1,\si_2} \nu[a,b] =\sum_p \nu[q] \exp{- i (a-b) q} \nn \\
\nu[a,b] &=& \ \frac{1 - \frac{n}{2} }{1- n}  \{ (\frac{n}{2} -1) \chi^{(1)}[a,a;b]+ \frac{n}{2} 
\ \chi ^{(2)}[a,a;b] -   \chi ^{(3)}[a,a;b] \}.\nn \\
&=& \frac{1}{1-\frac{n}{2}} \left\{ \right \langle \vec{S}_a\cdot \vec{S}_b \rangle + \frac{1}{4} ( \langle n_a n_b \rangle-n^2)  \}, \;\;\text{with} \;\tau_a=\tau_b \nn \\
 \nu[q] &=& -\frac{1}{2} (1-\frac{n}{2}) \chi_s[q] - \frac{3}{2} \ \frac{1- \frac{n}{2}}{1-n}  \ \chi_t[q]. \llabel{nu-1}
   \earray

Further on turning off the sources, we find from Appendix \ref{app3}
\barray
\Theta_{\si_1,\si_2}[r,s,m] & = &  {\si_1 \si_a} \mu_{\si_a,\si_b}[r,s]  \GL_{\si_b,\si_c}[s,k] \Gamma_{\sib_1, \sib_a}^{\si_c,\si_2}[k,m;r] 
 \llabel{theta_def_3}\nn \\
&\rightarrow& \delta_{\si_1,\si_2} G[s,k] \; \{ \Gamma
   ^{(2)}[k,m;r]-    \Gamma^{(3)}[k,m;r] \} \nn \\
&=& \delta_{\si_1,\si_2} G[s,k] \Gamma^{(p)}[k,m,r]. \llabel{def-Theta}
\earray
 As explained in Appendix \ref{app5}, the superscript $(p)$ stands for Cooper pairing channel and is a specific linear superposition 
of the singlet and triplet channels $\Gamma^{(p)}= \frac{1}{2}(\Gamma_{s}- 3 \Gamma_{t} ) $. Hence we may write the EOM as
\barray
 && - \GLI[i,j]  =    (\partial_{\tau_i} - \mu     + \Phi_i )  \delta[i,j]  \nn \\
&&    -t[i,j]  \  ((1-\frac{n}{2}) +  \nu[i,j])  - t[i,l]  \   G[l,k] \Gamma^{(p)}[k,j;i]   \nn \\
&& + \frac{1}{2} J[i,k] \  ( \nu[k,i] +   (1-\frac{n}{2}) ) \ \delta[i,j]   +  \frac{1}{2} J[i,k] \  G[i,l] \Gamma^{(p)}[l,j;k]
). \nn \\
 \llabel{eom_7}
\earray

Using the convention for Fourier transforms in Appendix \ref{app6},
and with  the Fourier space version of $\GLO$ Eq(\ref{gsplit})
\beq
\GLIO[k]= i \omega_k + \mu - (1-\frac{n}{2})(\varepsilon_k + \frac{1}{2} J_0).\llabel{g0-2}
\eeq
 we write the exact  Schwinger Dyson equation for the Greens function
\barray
\GL^{-1}[k]  &=&  \GLIO[k] - \self[k], \nn \\ 
\self[k]&=&   \sum_q \ (\varepsilon_{q+k} + \frac{1}{2}  J_q) \nu[q] + \frac{1}{2} \sum_q \left( \varepsilon_q + \frac{1}{2} J_{k-q} \right) \ \GL[q] 
\ \{ \ \Gamma_{s}[q,k]- 3 \ \Gamma_{t}[q,k] \}. \nn \\ \llabel{schwinger_dyson}
\earray  
 The self energy explicitly contains the electron dispersion $\varepsilon_q $ or  $t_{ij}$  
as a multiplicative factor here. This is quite unlike the standard case of the electron liquid, where the entire
dependence on the electronic dispersion is hidden in the form of the bare propagator $G_0$, and the self energy is a functional of only the (full) Greens function\cite{luttinger_ward,agd,rajagopal,mahan}.

In order to evaluate the first term of  $\self$ in Eq(\ref{schwinger_dyson}),  we need a few definitions at this point. Let us denote
\begin{align}
\chi_s[Q] &= \sum_q \chi_s[q,q+Q], 
 &\chi_{ \loc} = & \sum_Q \; \chi_s[Q]  = -\frac{n}{2}\ \frac{(1-n)}{(1-\frac{n}{2})^2},& \nn \\
 \varepsilon^s_{p} =& \frac{1}{\chi_{\loc}} \sum_Q \; \varepsilon_{p+Q}\  \chi_s[Q],  
& J^s_{p} =& \frac{1}{\chi_{\loc}} \sum_Q \; J_{p+Q}\  \chi_s[Q].  
 \llabel{energies-1} 
\end{align}
Similar definitions hold in the triplet channel.
In appendix \ref{app9}  we show that the local object $\chi_{\loc}$ is identical  for both triplet and singlet channels. 
  From Eqs(\ref{nu-1}, \ref{schwinger_dyson} and \ref{energies-1}), and further using the definition of the renormalized energy type variables
\barray
\ehat_k& = &  \sum_q \ \varepsilon_{q+k} \  \nu[q] =   \frac{3}{4} \frac{n}{(1-\frac{n}{2})}\ \varepsilon_k^t + \frac{1}{4} \frac{n (1-n)}{(1-\frac{n}{2})}\ \varepsilon_k^s         \nn \\
\Jhat_k& = &  \frac{1}{2} \sum_q \    J_{q+k} \ \ \nu[q]=  \frac{3}{4} \frac{n}{(1-\frac{n}{2})}\ J_k^t + \frac{1}{4} \frac{n (1-n)}{(1-\frac{n}{2})}\ J_k^s
\llabel{energies-2}
\earray
we find
\barray
\self[k]=  (\ehat_k + \frac{1}{2} \Jhat_0) + \frac{1}{2} \sum_q \left( \varepsilon_q + \frac{1}{2} J_{k-q} \right) \ \GL[q] \ \{ \ \Gamma_{s}[q,k]- 3 \ \Gamma_{t}[q,k] \}. \nn \\ \llabel{schwinger_dyson-2}
\earray  
By construction, the variables $J^s_{p}, J^t_{p},  \varepsilon^s_{p}, \varepsilon^t_{p} $ are weighted
averages of the bare objects $J_p, \varepsilon_p$ with momentum dependent but static weight factors, and hence it is reasonable to view them as
correlation adjusted versions of the bare dispersions. The precise relation between the bare dispersions and $\ehat_k, \Jhat_k$ is particularly simple 
when the hopping is only nearest neighbour; it involves the 
the spin spin and density density correlation functions at nearest neighbour distances,  and is given explicitly below in Eq(\ref{ehat-2}).
 Specializing to only nearest neighbour hoppings, we see from symmetry 
that the form of the band dispersion remains a simple tight binding one. Therefore we write a convenient notation of 
various objects for the simple cubic lattices:
\beq
x_{t, \ s}  \equiv  \frac{1}{\chi_\loc} \ \sum_q  \cos q_x \  \chi_{ t, \ s}[q], \llabel{xt-def}
\eeq
The  parameters $x_s$ and $x_t$ can be found in terms of equal time correlations by carrying out the frequency and spatial sums, and using Eqs(\ref{ups_chi},\ref{ups_chi_3}), we find
\begin{align}
 x_s& = \frac{\langle n_{\vecr_i} \ n_{\vecr_i+\vecn} \rangle -n^2}{n -n^2}, \;\;\text{and} & x_t&= \frac{4}{3 \ n } \
 \langle \vec{S}_{\vecr_i} \cdot \vec{S}_{\vecr_i+\vecn} \rangle,  \llabel{afmcf}
\end{align}
where $\vecn$ is a nearest neighbour vector.  In terms of these, we may write  
\begin{align}
  \varepsilon_k^t& =   x_t \ \varepsilon_k,  &\varepsilon_k^s& =   \varepsilon_s \ J_k,   & J_k^t& =   x_t \ J_k,   & J_k^s& =   x_s \ J_k.  
\llabel{jt-js}
\end{align}  
Thus we can express 
\begin{align}
\ehat_k &=  \frac{\varepsilon_k}{1-\frac{n}{2}} \left\{ \right \langle \vec{S}_{\vec{0}}\cdot \vec{S}_{\vecn} \rangle + \frac{1}{4} ( \langle n_{\vec{0}} \ n_{\vecn} \rangle-n^2)  \}, \nn \\
\Jhat_k& = \frac{J_k}{1-\frac{n}{2}} \left\{ \right \langle \vec{S}_{\vec{0}}\cdot \vec{S}_{\vecn} \rangle + \frac{1}{4} ( \langle n_{\vec{0}} \ n_{\vecn} \rangle-n^2)  \}, \llabel{ehat-2}
\end{align}
where $\vecn$ is a nearest neighbour vector and the correlations are at equal times.
Thus the nearest neighbour charge and spin correlations directly influence the effective band width.
The spin correlation  has its  largest magnitude near half filling, and  can be positive or negative  depending upon whether ferromagnetic, i.e. Nagaoka Thouless type correlations,  or the more usual antiferromagnetic correlations prevail.
Near half filling the density   term in Eq(\ref{ehat-2}) is suppressed, while the spin correlation  term survives and gives the dominant contribution.
This correction term to the energy dispersion $\ehat$ is of the same form as $\varepsilon$ but has the opposite sign and hence leads to an important
band narrowing via the correlation functions indicated in Eq(\ref{ehat-2})   in this theory.

Later we will find equations for the susceptibilities, so that these correlation functions can be found in terms of 
the vertices and the Greens functions self consistently. Successive approximations for the vertices will be formulated, with each
approximation    providing a   complete  calculational scheme.
We will find that the susceptibilities $\chi_{s,t}[Q]$ vanish  generically as $1-n$ near half filling, in a  similar fashion as  $\chi_{\loc}$, and hence these renormalized energies tend to a finite non zero limit at half filling.

Eq(\ref{schwinger_dyson-2}) is the  exact   Schwinger Dyson  equation for the $t$-$J$ model. This breakup of the inverse $\GL$ into $\GLO$  and a 
 self energy type object $\self$ is to some extent arbitrary, since there is no {\em a priori}  notion of an unperturbed Greens function.
Our breakup has the natural advantage that the resulting  bare vector vertices  are 
frequency independent and 
in agreement with  independent arguments (see next section where
the bare current vertex is obtained in Eq(\ref{kubo-2})). Further they satisfy the Ward identities as noted below  Eq(\ref{bare-vertex-2}).   
 The $\self$ object, after the standard analytic continuation
to the retarded self energy, provides a correction to the quasiparticle energies
at the poles of $\GL$ through its real part  and more importantly to the decay of the quasiparticles through its imaginary part. The self energy in our formulation
is developed in terms of the Greens function $\GL$ and is explicitly a functional of $\GL$. 
 However, and in contrast to the usual weakly interacting
Fermi system formulation\cite{pothoff} \cite{luttinger_ward,agd,nozieres}, it is {\em not a universal functional}, i.e. it does depend upon the bare
dispersion $t_{ij}, J_{ij}$ regardless of how we define the self energy.

Finally it is important to note that this  Schwinger Dyson equation  is well formed in the sense that the vertices $\Gamma$ are guaranteed to be well behaved (i.e. finite) for $\omega \rightarrow \infty$, since we have  avoided
the  linear $\omega$ dependence of $\Gamma$ that results, if one does not
extract the time dependent factor $\Delta[i]$  as in Eq(\ref{gauge_transform}). The equations for the  vertices are derived and  discussed  below.
\section{ Solution of the local Greens function \llabel{app4} }
We next consider the various local Greens functions and their interrelations. Start from
\barray
\G[j^-,j] &= & \Delta[j] \cdot \GL[j^-,j], \; \mbox{ with } \nn \\
\Delta[j]& = & \iden- \G^k[j^-,j], \; \mbox{ so that } \nn \\
\Delta[j]&=& \iden -\GL^k[j^-,j] \cdot \Delta^k[j] 
\earray
This equation is easy to solve when we iterate once,
\barray
\Delta[j] &=& \iden -\GL^k[j^-,j] + \GL^k[j^-,j] \cdot  \Delta[j] \cdot \GL[j^-,j], 
\earray
we see that the inhomogeneous term $\iden- \GL^k$ commutes with $\Delta$ iteratively, so that we may as well rewrite the  second term in the RHS as
$\Delta[j] \cdot \GL^k[j^-,j] \cdot  \GL[j^-,j]  = \det[G] \Delta[j]$, whereby the solution is noted in Eq(\ref{delta-sol-2});
\barray
\Delta[j]& =& \frac{1}{\gamma[j]} \left( \iden - G^k[j^-,j] \right), \nn  \mbox{ where }\\
\gamma[j] & = &1-\det[G[j^-,j]]. \llabel{eq61} 
\earray
We also require the inverse
\barray
\Din[j]& =& \frac{\gamma[j]}{\gamma1[j]} \left( \iden - G[j^-,j] \right), \nn  \mbox{ where }\\
\gamma1[j] & = &\det[\iden- G[j^-,j]]. \llabel{delta-sol-1}
\earray
These local objects are the dynamical analogs of the Gutzwiller projection factors \cite{Gutz}, and we note their physical meaning in terms of the bare (quasiparticle)  charges $n[i]$ ($n_{QP}[i]$)  and spin densities $\vec{S}[i]$ ($\vec{S}_{QP}[i]$) defined in Eqs(\ref{bare-density},\ref{transform-qp})
\begin{align}
\gamma[j]&= \frac{1-n[i]}{(1-\frac{1}{2}n[i])^2- \vec{S}[i]\cdot \vec{S}[i]} = 1 - \frac{1}{4} (n_{QP}[i])^2 + \vec{S}_{QP}[i]\cdot \vec{S}_{QP}[i] \nn \\
\gamma1[j]&= \frac{(1-n[i])^2}{(1-\frac{1}{2}n[i])^2- \vec{S}[i]\cdot \vec{S}[i]} = (1 - \frac{1}{2} n_{QP}[i])^2 - \vec{S}_{QP}[i]\cdot \vec{S}_{QP}[i]. \nn 
\end{align}
 
When we turn off the sources, the ECQL  state has the following local Greens functions
expressed in terms of the number density at each site per spin $n= N/( 2 N_{sites})$:
\begin{align}
\G[j^-,j]&= \frac{n}{2} 
&\GL[j^-,j]& =  \frac{\frac{n}{2}}{1-\frac{n}{2}} 
&\Delta[j]&=  (1 - \frac{n}{2}), \nn \\
\gamma[j]& =  \frac{1-n}{(1- \frac{n}{2})^2}
&\gamma1[j]&=  \frac{(1-n)^2}{(1- \frac{n}{2})^2}.&& \llabel{local-gamma-sourcefree}
\end{align}
The vanishing near half filling of $\gamma$ and $\gamma1$, as $(1-n)$ and $(1-n)^2$ has important consequences, and
leads to the hole density expansion reported below in Section \ref{nlha}.

\section{ Calculation of the Vertices \llabel{vertices}}
The vertices can be obtained from the general equations for the Greens function Eq(\ref{gsplit}, \ref{schwinger_dyson_self_energy}) as
\barray
\Gamma^{(1)}[i,j;k] & = & \delta[i,j] \ \delta[i,k] + \left(\frac{\delta}{\delta V_k^{\ua \ua}}\self_{\ua \ua}[i,j]\right)_{V \to 0} \nn \\
\Gamma^{(2)}[i,j;k] & = & \left( \frac{\delta}{\delta V_k^{\da \da}}\self_{\ua \ua}[i,j]\right)_{V \to 0} \nn \\
\Gamma^{(3)}[i,j;k]&=& \Gamma^{(1)}[i,j;k]- \Gamma^{(2)}[i,j;k]. \llabel{vertices-def}
\earray
Thus the bare vertices are simple, writing them in Fourier space we find
\begin{align}
\Gamma_s[p_1,p_2] &\to 1, & \Gamma_t[p_1,p_2]& \to 1. \llabel{bare-vertex} 
 \end{align}
The vertex corrections arise from the self energy part of Eq(\ref{vertices-def}), and contain several terms. We provide the  results of the  long calculation in  Appendix \ref{vertices_all}. There  all terms are retained, with the  proviso that the  higher order vertex,  obtained by differentiating the three point vertex are set to zero, i.e. $ \frac{\delta \Gamma}{\delta V} \to 0$.

While this complete solution is available  in the Appendix \ref{vertices_all}, it is crucial to understand the relative order of various terms, in order to make sensible
approximations. We next provide a set of calculations that give us such an understanding, we evaluate the derivatives of the 
basic elements $\Delta[i], \ \mu[i,j]$ and $\nu[i,j]$ occurring in the self energy, and show that these have an  explicit dependence upon the hole
density $\delta = 1-n$, measured from half filling. The terms can in fact be grouped in a formal expansion in the inverse hole density $\lambda= \frac{1}{1-n}$.
In addition there is an implicit dependence on $n$ in all terms,  and  certain terms vanish with $\delta$. Thus 
the final result for the vertex obtained by taking a product of the apparently divergent terms and the coefficients thereof
either vanish at half filling $n=1$ or remain finite. This gives us an organizing principle for grouping the contributions, the leading terms near half filling
consist of terms that remain finite at $n=1$, and one can throw out the rest, thereby giving us a {\em hole density expansion} that has been
conjectured earlier \cite{early_tJ_1}. 
 
Therefore taking the derivatives of the   Schwinger Dyson  equation with sources Eq(\ref{schwinger_dyson_self_energy}),
we  find  that the expansion of the vertices can be arranged as a formal power series in the {\em inverse hole density} $\lambda$,  in the form  $\Gamma = 1 + \Gamma_0 + \lambda \Gamma_1+ \lambda^2 \Gamma_2$. It is implied that the coefficients $\Gamma_j$
contain terms that  are either vanishing at half filling or are finite, and  rules for recognizing this implicit dependence are  provided later. In Appendix \ref{vertices_all} we list the  vertices $\Gamma_0,\Gamma_1,\Gamma_2$.

The existence of this expansion  is fortunate since the most interesting 
physical regime for a doped Mott insulator  is in the limit $n \to 1$. The factors of $\lambda$ arise from the dynamical  projection factors $\Delta[i], \ \mu[i,j]$ and $\nu[i,j]$ in the local Greens functions and the self energy.  We next  list the leading behaviour of the functional
 derivatives of the various matrix functions $\mu[i,j], \ \nu[i,j], \ \Delta[i]$   w.r.t. the source terms, so that results for  the vertices can be assembled together.
For this section, we will use spin diagonal sources $\W_m$ so that the matrices can all be taken as diagonal in spin space. We will need to use the Nozi\'eres relations $\chi^{(1)} \pm \chi^{(2)}= \chi_{s, \ t} $, in order to obtain the  complete the set of derivatives in the singlet and triplet channels. After providing 
the results, we will discuss their relative magnitudes, and also the implicit density dependence of the coefficients.

Let us begin with the less singular terms  $\Delta$ and $\mu$, where the functional derivative is explicitly  $ O(\frac{1}{\delta})$, and then progress to   $\nu$ which is $O(\frac{1}{\delta^2})$.  At some places, we will use the symbol $\sim$ to indicate that the density $n$ has been set at unity in all terms except the singular $\frac{1}{(1-n)^a}$ type terms.

\subsubsection{$\Delta[i]$ and its derivatives}
Let us recall for spin diagonal sources (Appendix \ref{app8})
\begin{align}
\Delta_{\ua \ua}[i] & =\frac{1}{\gamma[i]} \left( 1- G_{\da \da}[i^-,i] \right) &   \Delta_{\ua \ua}[i] \ \Bigr \rvert _{\W \rightarrow 0} = (1-\frac{n}{2}) \nn \\
\gamma[i]&= 1- \GL_{\ua \ua}[i^-,i] \ \GL_{\da \da}[i^-,i].  & \frac{\delta \gamma[i] }{\delta \W_m^{\si \si}} \ \Bigr \rvert _{\W \rightarrow 0} = - \frac{n}{2-n} \ \chi_s[i^-,i;m]. 
\end{align}
Therefore 
\begin{align}
\frac{\delta \Delta_{\ua \ua}[i] }{\delta \W_m^{\ua \ua}} \ \Bigr \rvert _{\W \rightarrow 0} &= \frac{(1-\frac{n}{2})^2}{(1-n)} 
\left[ \frac{n}{2} \chi_s[i^-,i;m]-\chi^{(2)}[i^-,i;m] 
\right]  & \sim   \frac{1}{ 8 \delta} & \ \chi_t[i^-,i;m] \nn \\
\frac{\delta \Delta_{\ua \ua}[i] }{\delta \W_m^{\da \da}} \ \Bigr \rvert _{\W \rightarrow 0} &= \frac{(1-\frac{n}{2})^2}{(1-n)} 
\left[ \frac{n}{2} \chi_s[i^-,i;m]-\chi^{(1)}[i^-,i;m] \right]  & \sim  - \frac{1}{ 8 \delta} & \ \chi_t[i^-,i;m] . \llabel{delta-derivatives}
\end{align}

\subsubsection{$\mu[i,k]$ and its derivatives}
Lets us recall $\mu[i,k]= \Din[i] \cdot \Delta[k]$, and  it may be expressed as
\beq
\mu_{\ua \ua}[i,k] = g_2[k,i] \ (1- \GL_{\ua \ua}[i^-,i]) (1- \GL_{\da \da}[k^-,k]),\llabel{mu-again}
\eeq
where $g_2[i,j]$ and a related object $g_1[i,j]$ needed in the next section  are given by
\begin{align}
g_1[i,j] & = \frac{\gamma[j]}{\gamma1[j] \ \gamma[i]^2},   & g_2[i,j]  = \frac{\gamma[j]}{\gamma1[j] \ \gamma[i]}, \llabel{g_1}\\
g_1[i,j] \ \Bigr \rvert _{\W \rightarrow 0}  & = \frac{(1-\frac{n}{2})^4}{(1-n)^3},  & g_2[i,j] \ \Bigr \rvert _{\W \rightarrow 0}   = \frac{(1-\frac{n}{2})^2}{(1-n)^2} , \llabel{g-1-lim}
\end{align}
where  we used  the definitions (recall Eqs(\ref{delta-sol-1}, \ref{delta-sol-2})).

We first note the derivatives of $g$
\begin{align}
\frac{\delta  g_1[i,j] }{\delta \W_m^{\si \si}}  \ \Bigr \rvert _{\W \rightarrow 0}  & = \frac{(1-\frac{n}{2})^5}{(1-n)^4}    \left[  {n} \chi_s[i^-,i;m] + (1- \frac{n}{2}) \chi_s[j,j;m]  \right] & \sim \frac{1}{64 \delta^4} \left[  2 \chi_s[i^-,i;m] +  \chi_s[j,j;m]  \right]  \\
 \frac{\delta  g_2[i,j]}{\delta \W_m^{\si \si}}  \ \Bigr \rvert _{\W \rightarrow 0}  &  = \frac{(1-\frac{n}{2})^3}{(1-n)^3} \left[  \frac{n}{2} \chi_s[i^-,i;m] + (1- \frac{n}{2}) \chi_s[j,j;m]  \right] & \sim \frac{1}{16 \delta^3} \left[   \chi_s[i^-,i;m] +  \chi_s[j,j;m]  \right].
\end{align}
Notice that the derivatives of $g_1$ and $g_2$  are independent of the spin index $\si$.
We next compute the derivatives of $\mu[i,k]$ with spin  diagonal sources, by using the above equations and the derivatives of the second and third factors of Eq(\ref{mu-again}) :
\begin{align}
\frac{\mu_{\ua \ua}[i,k]  }{\delta \W_m^{\ua \ua}}  \ \Bigr \rvert _{\W \rightarrow 0}   
&=  \frac{(1-\frac{n}{2})}{2 (1-n)} \left[   \chi_t[k^-,k;m]  -  \chi_t[i^-,i;m] + (1-n) \{\chi_s[i^-,i;m] - \chi_s[k^-,k;m] \}    \right] \nn \\
& \sim \frac{1}{4 \delta} \left(  \  \chi_t[k^-,k;m]  - \chi_t[i^-,i;m] \  \right),  \;\;\mbox{ and similarly} \\
\frac{\mu_{\ua \ua}[i,k]  }{\delta \W_m^{\da \da}}  \ \Bigr \rvert _{\W \rightarrow 0} & = \frac{(1-\frac{n}{2})}{2 (1-n)} 
\left[      \chi_t[i^-,i;m] - \chi_t[k^-,k;m] + (1-n) \{\chi_s[i^-,i;m] - \chi_s[k^-,k;m] \}    \right]
\nn \\
& \sim 
 \frac{1}{4 \delta} \left( \  \chi_t[i^-,i;m] -  \chi_t[k^-,k;m]    \ \right).\llabel{mu-derivatives}
\end{align}

\subsubsection{$\nu[i,k]$ and its derivatives}
Using the functions $g_1 \ g_2$ in Eq(\ref{g_1}), We first write $\nu$   as
\barray
\nu[i,k]& =& \nu_1[i,k]+ \nu_2[i,k] \nn \\
\nu_1[i,k] & =& g_1[i,k] \  \bar{\nu}[i,k],  \;\;\;\mbox{and}\;\;\; \nu_2[i,k] =g_2[i,k] \ \bar{\bar{\nu}}[i,k],
\earray
and assuming spin diagonal sources, we find from Eq(\ref{nu_def});
\barray
\bar{\nu}_{\ua \ua} [i,k] & = & (1- \GL_{\ua \ua}[i^-,i]) (1- \GL_{\ua \ua}[k^-,k]) 
\ \left\{ \GL_{\ua \ua}[i^-,i] \  \chi^{(1)}[i^-,i;k] + \GL_{\da \da}[i^-,i] \ \chi^{(2)}[i^-,i;k]
\right\}, \nn \llabel{nubar} \\
\bar{\bar{\nu}} _{\ua \ua} [i,k] & = & - (1- \GL_{\ua \ua}[k^-,k]) \ \  \chi^{(1)}[i^-,i;k] - (1- \GL_{\da \da}[k^-,k]) \ \  \chi^{(3)}[i^-,i;k].  \llabel{nubarbar} 
\earray
In order to obtain the leading order terms in inverse hole density, we note that the factors $1-\GL$ generate factors of $1-n$ which lower the order of the term, unless we differentiate these terms first. Hence to leading order, the calculation is simply done by ignoring the source dependence of the $\chi's$
 in Eq(\ref{nubarbar}), and differentiating the factors of $1-\GL$. We  thus obtain the leading derivatives of the two objects $ \nu_1,  \nu_2$ as follows
\small
 \begin{align}
 \frac{\delta  {\nu_1}_{\ua \ua} \ [i,k] }{\delta \W_m^{\ua \ua}}  \ \Bigr \rvert _{\W \rightarrow 0}  & = \frac{1}{16 \delta^2} \chi _{s}[i^-,i;k] \left(\chi
   _{s}[i^-,i;m]+\chi _{t}[i^-,i;m]-\chi
   _{t}[k^-,k;m]\right) \nn  \\
 \frac{\delta  {\nu_1}_{\ua \ua} \ [i,k] }{\delta \W_m^{\da \da}}  \ \Bigr \rvert _{\W \rightarrow 0}& =  \frac{1}{16 \delta^2} \chi _{s}[i^-,i;k] 
 \left(\chi_{s}[i^-,i;m] - \chi _{t}[i^-,i;m] + \chi_{t}[k^-,k;m]\right) \nn \\
 \frac{\delta  {\nu_2}_{\ua \ua} \ [i,k] }{\delta \W_m^{\ua \ua}}  \ \Bigr \rvert _{\W \rightarrow 0}  &= \frac{1}{16 \delta^2} \left\{ \ \chi _s[i^-,i;k] 
 \left(\chi _t[k^-,k;m]-\chi_s[i^-,i;m] \right)-\chi _t[i^-,i;k] \left(3 \chi _s[i^-,i;m]+\chi_t[k^-,k;m] \right)\ \right\}  \nn  \\
 \frac{\delta  {\nu_2}_{\ua \ua} \ [i,k] }{\delta \W_m^{\da \da}}  \ \Bigr \rvert _{\W \rightarrow 0}  &=
 \frac{1}{16 \delta^2 } \left\{ \ \chi _t[i^-,i;k] \left(\chi _t[k^-,k;m]-3 \chi
   _s[i^-,i;m]\right)-\chi _s[i^-,i;k] \left(\chi _s[i^-,i;m]+\chi
   _t[k^-,k;m] \right)\ \right\} \llabel{nu-derivatives}
   \end{align} 
\normalsize
 Using the Nozi\`eres relations we can combine these and write the final answers in the singlet and triplet channels
 \barray
 \frac{\delta  {\nu} \ [i,k] }{\delta \W_m } \ \Bigr \rvert _{\W \rightarrow 0}^{singlet}&=&  -\frac{3}{8 \delta^2} \chi _s[i^-,i;m] \chi _t[i^-,i;k] \nn \\
 \frac{\delta  {\nu} \ [i,k] }{\delta \W_m } \ \Bigr \rvert _{\W \rightarrow 0}^{triplet}&=& \frac{1}{8 \delta^2} \left\{ \ \chi _s[i^-,i;k] \chi _t[i^-,i;m]-\chi _t[i^-,i;k]
   \chi_t[k^-,k;m]\ \right\}. \llabel{nu-derivatives-2}
 \earray

Let us now consider expressions Eq(\ref{delta-derivatives}, \ref{mu-derivatives},\ref{nu-derivatives}), where we find terms with $\lambda, \lambda^2$ explicitly appearing ( where $\lambda = \frac{1}{\delta}$). The final answer  of each term is either finite or vanishes, and to see this we must recognize the implicit dependence on $\delta$ of the coefficients. The two point susceptibilities $\chi_{s ,\ t}[i,i^-;j]$, with arbitrary arguments are  seen to vanish linearly  as $\delta \to 0$ since these involve a particle and a hole in the quasiparticle band, which becomes completely filled at half filling. By the same token all three point susceptibilities also  vanish linearly as $\delta \to 0$. We should also remember that the vertices themselves are non vanishing as $\delta \to 0$.
This behaviour is readily  confirmed by a few low order calculations. As a useful example of this organization, we next write all the terms in the vertices that have an explicit quadratic dependence on $\lambda$. These are obtained by  assembling the  explicitly  $O(\lambda^2)$ terms in Eq(\ref{delta-derivatives}, \ref{mu-derivatives},\ref{nu-derivatives}), giving us
\barray
\Gamma_s[i,j;m]&=& \delta[i,m] \delta[j,m] + \frac{3}{8} \; \lambda^2 \;\{ t[i,j] \; \chi_s[i,i;m] \; \chi_t[i,i;j] - \delta[i,j]\; \frac{1}{2} J[i,k] \chi_s[k,k;m]\; \chi_t[k,k;i]\} \nn \\
\Gamma_t[i,j;m]&=&  \delta[i,m] \delta[j,m]  -  \frac{1}{8} \; \lambda^2 \;   t[i,j] \; \left\{ \chi_s[i,i;j]\;\chi_t[i,i;m] - \; \chi_t[i,i;j]\;\chi_t[j,j;m] \right\}  \nn \\
&& -  \frac{1}{16} \; \lambda^2 \;\delta[i,j]\; J[i,k] \; \left\{  \chi_t[k,k;m](3 \; \chi_t[k,k;i]\ -\chi_s[k,k;i]) - 2 \; \chi_t[i,i;m]\;\chi_t[k,k;i] \right\}\nn \\ \llabel{nlha_1}
\earray
The vertex corrections are clearly non zero as $\delta \to 0$ since the factor $\lambda^2$ is compensated by two vanishing susceptibilities. These can be recognized from the detailed list in the Appendix\ref{vertices} Eqs(\ref{app-gamma-s},\ref{app-gamma-t}) as arising from $\Gamma[i,j;m]_2$ and $\Gamma[i,j;m]_5$ in both singlet and triplet channels.
We will see next that these vertices satisfy current conservation and hence are studied further in Sec\ref{nlha}. We  emphasize that these do not exhaust 
the set of terms that are finite as $\delta \to 0$, and constitute  a convenient  subset of the  surviving terms. 

\section{Electrical Conductivity, Current conservation and  Ward identities \llabel{sec-6}}

We next set up the calculation of the electrical conductivity for the ECQL. We are interested in  deriving 
 a  useful  Kubo type expression  for conductivity  
in the \tJ model, in terms of the  appropriate Greens functions developed here.
 We will also  establish  Ward identities\cite{ward,takahashi,ma-bealmonod}  that relate the current and charge vertices, and find explicit expressions for the
current vertex. 
 These Ward identities  are important relations since they constrain the possible approximations one makes, and are
  necessary satisfy gauge and rotation invariance of the final results.
The charge and number densities and currents are trivially related by a factor of $q_e$ the electronic charge, and so we will
work with the number densities only. The number density $ n[\vecr \ \tau] =  \X{\vecr }{\si \si}( \tau) $ 
 has an  associated  current density   given by
\begin{align}
 \jc(\vecr)&\equiv \frac{1}{2}\sum_{\vecn} \jc(r+ \frac{1}{2}\vecn ),
 &\jc(r+ \frac{1}{2}\vecn)&\equiv i t_{\vecn} \ \vecn \  ( \X{\vecr+\vecn}{\si 0} \X{\vecr}{0 \si} -  \X{\vecr}{\si 0} \X{\vecr+\vecn}{0 \si}) 
  \llabel{currents} \\
\end{align}
where $\vecn$ is the set of  unit vectors connecting a site to its neighbours,  $\sum_\si$ is implied in all terms. Similar expressions can be written down also for the
spin density conservation, but since the \tJ model involves terms that flip the spin explicitly, the form of the resulting equations differ from the charge 
Ward identities\footnote{On the other hand, for the Hubbard model,  it is easy to show that the triplet vertices satisfy  finite versions  of the Ward identity analogous to Eq(\ref{ward-3}, \ref{ward-4}), since the spin density commutes with the interaction terms. } . We will be satisfied with using the Nozi\`eres relations to verify compliance with rotation invariance; these are necessary and most often sufficient conditions for giving us an invariant theory. The lattice version of divergence of these currents  is written  in terms of the $\ttau[\vecr]$ densities as
\begin{align}
(\vec{\nabla}_{\vecr} \cdot \jc(\vecr))& \equiv -i \ \ttau[\vecr ] =  \sum_{\vecn} \vecn \cdot \jc(\vecr+ \frac{1}{2} \vecn ),
 & \ttau[\vecr ]&= \sum_{\vecn \si} t_{\vecn} \ (    \X{\vecr}{\si 0} \X{\vecr+\vecn}{0 \si} - \X{\vecr+\vecn}{\si 0} \X{\vecr}{0 \si}). 
  \llabel{ttau}
 \end{align}
The total current operator can be written
\beq
\vec{J} = \sum_r \jc(r)= i \sum_{\vecr,\vecn}\ \vecn  \  t_{\vecn} \ \X{\vecr+\vecn}{\si 0} \X{\vecr}{0 \si}, \llabel{currents-2}
\eeq from Eq(\ref{currents}), and in case of external magnetic fields, we add a suitable Peierls phase factor to the currents. 
This expression for $\vec{J}$ is necessary for the calculation of the frequency dependent electrical  conductivity tensor   \cite{agd,mahan} 
of the \tJ model with $\Omega_n = 2 \pi k_B T n$
\barray
\sigma_{\alpha \beta}(i \Omega_n)& = & \frac{1 }{ N_S \ \Omega_n}\left\{ \ \langle \hat{\bf T}_{\alpha \beta} \rangle \  + \Pi_{\alpha \beta}(i \Omega_n) \right\} \nn \\ 
\Pi_{\alpha \beta}(i \Omega_n)& = & q_e^2 \int_0^\beta \ d \tau \ e^{i \Omega_n \tau}
 \ \langle \ T_\tau  J^\alpha(\tau) \ J^\beta  \ \rangle , \llabel{kubo}
\earray 
where we set the lattice constant $a_0$ and $\hbar$ to unity.
Here
$\hat{\bf {T}}_{\alpha \beta}$ is the diamagnetic part of the response related to the plasma sum rule \cite{shastry_review}, and may be evaluated as an equal time correlation.  As explained in standard texts\cite{agd,mahan}, this expression should be 
analytically continued as $ i \Omega_n \rightarrow \omega + i 0^+$, in order to obtain the retarded physical conductivity.
Our task is now to express $\Pi(i \Omega)$ in terms of the Greens functions in a manner that satisfies current conservation. 

Since we are primarily interested in  electromagnetic response, e.g. the  optical conductivity, the condition $Q a_0 \ll 1$ is satisfied. Hence
 we need long wave length response where the lattice structure is not crucial for the current conservation laws. However,
the calculation is most effectively performed by first noticing that one can find  a  lattice version of the
 current conservation law Eq(\ref{cons-laws}), valid at all length scales as detailed below. This more powerful conservation law\cite{takahashi} reduces to standard electrical conservation law $\partial_t \rho(r,t) + \nabla \cdot \vec{J}(r,t)=0 $ at long wave lengths. Our strategy 
is to derive the exact consequences of the lattice version of the  conservation law, and then to take the long wave length limit to obtain the 
electromagnetic vertices.

 Using the Heisenberg equations of motion in imaginary time, it is easy to establish the  exact  conservation laws of 
 number density    $\partial_\tau n[\vecr \ \tau]=  i (\vec{\nabla}_{\vecr} \cdot \jc(\vecr))$ compactly in operator form:
 \barray
 \partial_\tau n[\vecr \ \tau]&=& \ttau[\vecr \ \tau] , \llabel{cons-laws}
 \earray
where $\ttau$ is defined above in Eq(\ref{ttau}). The gauge   invariance of the \tJ  Hamiltonian is ultimately responsible for 
these conservation laws. In order to implement the conservation laws Eq(\ref{cons-laws}), we first establish the equation for the Greens functions as 
\beq
\partial_{\tau_m} \lll \X{i}{0\si_i} \X{f}{\si_f 0} \ n[m] \rrr -  \lll \X{i}{0\si_i} \X{f}{\si_f 0} \ \ttau[m] \rrr =
(\delta[i,m]-\delta[f,m]) \G_{\si_i \si_f}[i,f], \llabel{ward-1}
\eeq 
where the  terms on the RHS arise from the discontinuities of 
 time ordering implied in the Greens functions\cite{ward}. The two Greens functions in 
Eq(\ref{ward-1}) can be 
obtained from the action in Eq(\ref{action}) by adding terms
\beq
{\cal A} \rightarrow {\cal A} + \sum_{\vecr_m} \int_0^\beta  \ d\tau_m \ \left( u[m] \ n[m] + v[m] \ \ttau[m] \right), \llabel{action-2}
\eeq
where it can be seen that the effect of $u[m]$ is precisely that of  $\V^{\si \si}_m$ after summing over $\si$ and thus gives the singlet
susceptibilities and vertices. Our next task is to rewrite Eq(\ref{ward-1}) in terms of $\GL$. Towards this we recall that $\G[i,f] = \Delta[i] \cdot \GL[i,f]$,
and note that 
$$\left\{ \partial_{\tau_m}  \frac{\delta}{\delta u[m] }  - \frac{\delta}{\delta v[m] }  \right\} \Delta[i]=0, $$
on using the conservation law Eq(\ref{cons-laws}). Putting these together, 
 we may thus rewrite 
Eq(\ref{ward-1})  in matrix form  as
\barray
\left\{ \partial_{\tau_m}  \frac{\delta}{\delta u[m] }  - \frac{\delta}{\delta v[m] }  \right\}  \GL[i,f] & = &
(\delta[i,m]-\delta[f,m]) \  \GL[i,f],  \nn \\
\left\{ \partial_{\tau_m}  \frac{\delta}{\delta u[m] }  - \frac{\delta}{\delta v[m] }  \right\}  \GLI[i,f] & = &
(\delta[i,m]-\delta[f,m]) \  \GLI[i,f],  \nn \\
\llabel{ward-2}
\earray
where the disconnected parts cancel identically upon using the conservation law Eq(\ref{cons-laws}). The second form   is specially useful
for checking the compliance of conservation laws, it is the most  compact statement of  the Ward identity.

We define new ``$\ttau$- susceptibilities''  and   vertices 
\begin{align}
\chi_s^{\ttau}[i,j;m]& = \frac{\delta}{\delta v[m]} \GL[i,j],
& \Gamma_s^{\ttau}[i,j;m]& = - \frac{\delta}{\delta v[m]} \GLI[i,j]. \llabel{vertex-current}
\end{align}
The subscripts $s$ are to denote the singlet, i.e. charge  nature of these currents. Below we obtain the electromagnetic vertices from these by a limiting process.
These objects  satisfy the real space  Ward identities
\barray
\left\{ \partial_{\tau_m}  \chi_s[i,f;m]  - \chi_s^{\ttau}[i,f;m]  \right\}  & = &
(\delta[i,m]-\delta[f,m]) \  \GL[i,f], \;\;\;\; \text{or} \nn \\
\left\{ \partial_{\tau_m}  \Gamma_s[i,f;m]  - \Gamma_s^{\ttau}[i,f;m]  \right\}  & = &
(\delta[f,m]-\delta[i,m]) \  \GLI[i,f].  \nn \\
\llabel{ward-3}
\earray
Fourier transforming these we obtain the  momentum space Ward identities
\barray
i (\omega_{p_1}- \omega_{p_2}) \   \chi_s[p_1,p_2] - \chi^{\ttau}_s[p_1,p_2] & = & \GL[p_2]-\GL[p_1] \nn \\
i (\omega_{p_1}- \omega_{p_2}) \ \Gamma_s[p_1,p_2] - \Gamma^{\ttau}_s[p_1,p_2] & = & \GLI[p_1]- \GLI[p_2]. \llabel{ward-4}
\earray
We next evaluate the $\ttau$ vertex Eq(\ref{vertex-current}). The effect of $v[m]$
is most easily seen as a modification of the kinetic energy, since the $\ttau$ operator in Eq(\ref{ttau}) resembles the kinetic energy closely.
In fact we can write
\beq
\sum_m v[m] \tau[m] = \sum_{m,k} t_{m,k} (v[m]-v[k]) \X{m}{\si 0} \X{k}{0 \si}, \llabel{extrav}  
\eeq
and recalling the definition of the Hamiltonian  Eq(\ref{hamiltonian}), we can see that 
the effect of adding this term in  the relevant  time domain equations is to replace  $t[m,k] \rightarrow t[m,k](1+v[k]-v[m])$. With this observation, we may write 
the Greens function equation Eq(\ref{gsplit},\ref{schwinger_dyson_self_energy}) with these added sources. We again  split the Greens function as in Eq(\ref{gsplit})  
\barray
  \GL^{-1}[i,j] & = &  \GLIO[i,j] - \self[i,j],   \;\;\text{where}   \nn \\
 - \GLIO[i,j]  & \equiv &   (\partial_{\tau_i} - \mu   +  \W_i    +u[i] + \frac{1}{2} J[i,k] (1-\frac{n}{2}))  \delta[i,j]      \nn \\
&&  -t[i,j](1+v[j]-v[i])    (1-\frac{n}{2}),
 \llabel{gsplit_2}
\earray
and the self energy is given by
\barray
  \self[i,j] & = &    \Phi_i   \delta[i,j] -t[i,j](1+v[j]-v[i])  \  (\Delta[j] +  \nu[i,j] -(1- \frac{n}{2}))     \nn \\
&&    - t[i,l](1+v[l]-v[i])  \   \Theta[i,l,j] +  \frac{1}{2} J[i,k] \ \mu[i,k] \cdot \Theta[k,i,j] \nn \\
& & + \frac{1}{2} J[i,k] \  \left( \mu[i,k]\cdot \nu[k,i] +   \ \mu[i,k] \cdot \Delta[i] -(1-\frac{n}{2})\right) \ \delta[i,j]. \nn \\
&&  \llabel{schwinger_dyson_self_energy_added} \nn \\
\earray

The charge and current vertices follow from taking the derivatives of the above equation. All vertices are naturally split into bare and correction terms
\beq 
\Gamma = \gamma + \hat{\Gamma}, \llabel{vertex-split}
\eeq
where $\gamma \equiv - \frac{\delta}{\delta u}\GLIO $ and $\hat{\Gamma} =  \frac{\delta}{\delta u}\self $.
It follows from Eq(\ref{schwinger_dyson_self_energy_added}) that
\barray
\gamma_s[i,j;m] &=&  \delta[i,m]  \ \delta[j,m], \nn \\
\gamma_s^{\ttau}[i,j;m]&=& -t[i,j](\delta[j,m]-\delta[i,m])(1- \frac{n}{2}).   \nn \\
 \llabel{bare-vertex-1}
\earray
In Fourier space we find
\barray
\gamma_s[p_1,p_2] &=& 1, \nn \\
\gamma_s^{\ttau}[p_1,p_2]&=&  (\varepsilon_{ p_1} - \varepsilon_{ p_2} )  (1- \frac{n}{2}). \llabel{bare-vertex-2}
\earray
We observe that the bare vertices with Eq(\ref{g0-2}) satisfy the Ward identity Eq(\ref{ward-4}).

In order to make contact with our calculation of the  Schwinger Dyson vertices, we consider the susceptibility
\barray
\ups_s^{\ttau}[i,j;m]& \equiv & \frac{\delta}{\delta v[m]}\G_{\si \si}[i,j] =  \lll \X{i}{0 \si} \X{j}{\si 0} \ \ttau[m] \rrr - \lll \X{i}{0 \si} \X{j}{\si 0}  \rrr  \  \lll  \ttau[m] \rrr \nn \\
\ups_s^{\ttau}[p_1,p_2]&=& \sum_{i,j,m} e^{-i p_1 (i-m) - i p_2 (m-j)} \ups_s^{\ttau}[i,j;m] \nn \\
&=& \sum_{i,j,m,\vecn, \si' } e^{-i p_1 (i-m) - i p_2 (m-j)} \ t_{\vecn} \ (1- e^{ i \vecn \cdot (\vec{p}_1-\vec{p}_2)} ) \{\lll 
\X{i}{0 \si} \X{j}{\si 0} \X{m}{\si' 0} \X{m+ \eta}{0 \si'}
\rrr - \text{disc} \},\nn \\
\earray
where ``disc'' stands for the disconnected part and while $\si$ is fixed, $\si'$  is  summed over. In the second line onwards we turn off the sources,
hence  we may  ignore the ``disc'' pieces due to parity. We should  also note that $\eta$ is a  vector without a  time component,
so that the two  factors $\X{m}{} \X{m+\eta}{}$ are at the same time $\tau_m$.
We next use the definition of the current from  Eq(\ref{currents-2}), and take a specific limit:
\barray
&& - \frac{\partial}{\partial p'^\alpha}  \ups_s^{\ttau}[\vec{p} \ \omega_p,\ \vec{p'} \ \omega_p + \Omega_n]/_{p' \to p}  =  \nn \\
&& \sum_{ \vec{r}_i, \vec{r}_j} \int_0^\beta \int_0^\beta d(\tau_i- \tau_m) \ d(\tau_m - \tau_j) e^{ i \omega_p(\tau_i-\tau_j) + i \Omega_n (\tau_m -\tau_j) - i \vec{p} \cdot (\vec{r}_i-\vec{r}_j)} \lll \X{i}{0 \si}(\tau_i) 
\X{j}{\si 0}(\tau_j) J^{\alpha}(\tau_m)\rrr.  \nn\\
\earray
Convoluting with the velocity 
\beq 
v_p^\beta = \frac{\partial}{\partial p^\beta}\varepsilon_p = i \sum_{\vecn} \eta^\beta \ t_{\vecn} \ e^{ - i \vec{p} \cdot \vecn}, \llabel{velocity}
\eeq
we see that 
\barray
2 \sum_{p } e^{i \omega_p 0^+} v_p^\beta \ \left( \frac{\partial}{\partial p'^\alpha} \  \ups_s^{\ttau}[\vec{p} \ \omega_p,\ \vec{p'}
 \ \omega_p + \Omega_n]\right)_{p' \to p} & =& 
\int_0^\beta \ d (\tau_m-\tau_i) \ e^{i \Omega_n (\tau_m -\tau_i)} 
\lll J^\alpha(\tau_m) J^\beta(\tau_i) \rrr,\nn \\
&=& \Pi_{\alpha \beta}( i \Omega_n), \llabel{kubo-1}
\earray
where the factor of $2$ in front is from the spin summation. We note that the response functions $\ups$ can be related to the derivatives of $\GL$ in a straightforward way, as in Appendix \ref{app9}.
We obtain
\barray
\ups_s^{\ttau}[p_1,p_2]+ \ups_s^{\ttau} [p_2-p_1] \G[p_2] &=& (1-\frac{n}{2})\  \chi_s^{\ttau}[p_1,p_2] \nn \\
\ups_s^{\ttau} [p_2-p_1] & = & (1-\frac{n}{2})^2 \ \chi_s^{\ttau} [p_2-p_1].
\earray
 For notational convenience, in the following we write $p'=p+Q$, with $Q= \{ \vec{Q}, i \Omega_n \}$.  Thus we can rewrite Eq(\ref{kubo-1}) as
\barray
\Pi_{\alpha \beta}( i \Omega_n) & = & 2 (1- \frac{n}{2}) \ \sum_{p } e^{i \omega_p 0^+} v_p^\beta \ \left( 
\frac{\partial}{\partial Q^\alpha} \left\{ \chi_s^{\ttau}[p,p+Q] - (1- \frac{n}{2}) \chi_s^{\ttau}[Q] \GL[p+Q] \right \} \right)_{\vec{Q} \to 0},\nn \\ \llabel{kubo-2}
\earray
where the bare current vertex  $(1- \frac{n}{2}) v_p^\beta$ is  a consequence of  the definition of the current Eq(\ref{currents-2}).
We make the  implicit   assumption that $Q$ in the vector vertex $\Gamma_s^\alpha[p,p+Q]$ always satisfies $\vec{Q}a_0 \ll 1$, and all term of $O((Q^\alpha)^2)$ are thrown out,
 i.e. we are in the electromagnetic regime.
Observing from the bare vertex Eq(\ref{bare-vertex-2})that $\gamma_s^{\ttau}[p,p+Q]$,
 and hence the full vertex $\Gamma_s^{\ttau}[p,p+Q]$ vanishes as $\vec{Q} \to 0 $, and its derivative is odd under  parity, we conclude 
that the second term in Eq(\ref{kubo-2}) can be dropped. We may then define the vector (electromagnetic) vertex (with $\vec{Q} a_0 \ll 1$)
\barray
\ \Gamma_s^\alpha[p,p+Q]&=& - \frac{\partial}{\partial Q^\alpha} \Gamma_s^{\ttau}[p,p+Q] \nn \\
\chi_s^\alpha[p,p+Q]&=& \GL[p] \Gamma_s^\alpha[p,p+Q] \GL[p+Q],  \llabel{current-vertex}
\earray
with the bare vector  vertex $\vec{\gamma}$ as  
\barray
\vec{\gamma}_s[p,p + Q] &  = &  -  \left(\frac{\partial}{\partial \vec{Q}}\gamma_s^{\ttau}[p,p+Q]\right)_{\vec{Q} \to 0} =  \vec{v}_p (1-\frac{n}{2}),
 \llabel{bare-vertex-3}
\earray
so that
\beq
\Pi_{\alpha \beta}( Q )  = - \ 2 \ \sum_{p } \ e^{i \omega_p 0^+} \ \gamma_s^\beta[p,p+Q] \ \GL[p] \ \Gamma_s^\alpha[p,p+Q] \ \GL[p+Q]. \llabel{kubo-3}
\eeq
 The Ward identity Eq(\ref{ward-4}) relevant for electromagnetic response now reads
\beq
i \Omega_n \Gamma_s[p,p+Q] - \vec{Q} \cdot \vec{\Gamma}_s[p,p+Q] = \GLI[p+Q]- \GLI[p]. \llabel{ward-em}
\eeq
In practice, this infinitesimal form of the Ward identity Eq(\ref{ward-em}) is more easy to implement and also sufficient for the gauge invariance of the final theory, as compared with the finite form Eq(\ref{ward-4}). 

Returning to the Eq(\ref{vertex-split}), the equations for the vertex corrections $\hat{\Gamma}$ follow from taking derivatives of the self energy Eq(\ref{schwinger_dyson_self_energy_added}). We may write the charge vertex simply as
\beq
\hat{\Gamma}_s[i,j;m]= \left( \frac{\delta}{\delta u[m]} \self_{\si \si}[i,j]\right)_{u\to 0, v \to 0} =
\sum_{\si'} \left( \frac{\delta}{\delta \W_m^{\si' \si'}} \self_{\si \si}[i,j]\right)_{\W \to 0}
,
\llabel{density-vertex-1}
\eeq
with the spin index $\si$ fixed at  say $ \ua$. Here the current type sources $v[m]$ are irrelevant and could be set equal to zero right away in Eq(\ref{schwinger_dyson_self_energy_added}). The  second identity in Eq(\ref{density-vertex-1}) is a consequence of the rotation invariance
relations of Nozi\`eres
as discussed in Appendix \ref{app5}.

 In order to obtain the current type vertex corrections, we get two sets of contributions from Eq(\ref{schwinger_dyson_self_energy_added}).
We  write with say  a fixed $\si$, let us say  $\si= \ua$;
\barray
\hat{\Gamma}_s^{\ttau}[i,j;m] & = &  \hat{\Gamma}_{s \ a}^{\ttau}[i,j;m] + \hat{\Gamma}_{s \ b}^{\ttau}[i,j;m] \nn \\ 
\hat{\Gamma}_{s \ a}^{\ttau}[i,j;m] & = & \left( \frac{\delta}{\delta v[m]} \self'_{\si \si}[i,j]\right)_{u\to 0, v \to 0},\nn \\
\hat{\Gamma}_{s \ b}^{\ttau}[i,j;m] & = &(\delta[i,m]-\delta[j,m]) t[i,j] \nu_{\si \si}[i,j] + (\delta[l,m]-\delta[i,m]) t[i,l] \GL[l,k] \ \Gamma^{(p)}[k,j;i]. \nn \\
\llabel{current-vertex-1}
\earray
 The set of terms $ \hat{\Gamma}_{s \ b}^{\ttau}[i,j;m]  $ arise from the explicit factors of $v[m]$ in Eq(\ref{schwinger_dyson_self_energy_added}).
Here $\self'$ represents all the terms of Eq(\ref{schwinger_dyson_self_energy_added}), but with the explicit factors of $v[m]$ omitted. 
Thus the  term $ \hat{\Gamma}_{s \ a}^{\ttau}[i,j;m]  $ arises from differentiating
the rest of the $v[m]$ dependence,  and originate in  
 exactly the same terms that contribute to Eq(\ref{density-vertex-1}). The only difference is that after taking the derivative, 
 a current vertex replaces  the charge vertex.  Therefore one can obtain the equations satisfied by  
 $ \hat{\Gamma}_{s \ a}^{\ttau}[i,j;m]  $ from those obeyed by
$\hat{\Gamma}_s[i,j;m]$,  in a simple way. In the RHS,  we  
simply replace  terms indexed by the external  symbol $m$, e.g.  ${\Gamma}_s[p,q;m]$   by ${\Gamma}_{s }^{\ttau}[p,q;m]$ for any $p,q$.
The same separation is also carried out for the electromagnetic vertex $\Gamma_s^\alpha$, 
and may be obtained by taking the limit as in Eq(\ref{current-vertex}).
We note the vertices $\hat{\Gamma}_{s \ b}$ in Fourier space (with a fixed $\si$)
\barray
\hat{\Gamma}^{\ttau}_{s \ b}[p_1,p_2]&=& \sum_q (\varepsilon_{q+p_1}-\varepsilon_{q+p_2} )\ \nu_{\si \si}[q] + 
\sum_q (\varepsilon_{q+p_1-p_2}-\varepsilon_{q} ) \ G[q] \ \Gamma^{(p)}[q,p_2] \nn \\
\vec{\Gamma}_{s \ b}[p,p+Q]&=& \sum_q \vec{v}_{p+q} \  \nu_{\si \si}[q] + \sum_q \ \vec{v}_{q} \  G[q] \ \Gamma^{(p)}[q,p+Q]. \llabel{vector-vertex}
\earray

We summarize the requirements for a ``conserving approximation''\cite{kadanoff_baym} for the ECQL, it consists of a
suitable approximation for the self energy Eq(\ref{schwinger_dyson},\ref{schwinger_dyson-2}), the  density vertex Eq(\ref{density-vertex-1}) and the  
current vertex Eq(\ref{current-vertex-1},\ref{vector-vertex}). These must be consistent, i.e. satisfy the Ward identity Eq(\ref{ward-em})  so that the  response is gauge   invariant.
\section{General properties of the Greens Function and  Successive Approximations \llabel{sec-7} }
We next turn to computing the Greens functions from the above theory, after making a series of approximations. Let us summarize the various objects of interest.
 Our interest is initially in the positive 
definite 
spectral function $\rho_G[q,\nu]$, and the occupation of the $k$ state $m_k$  defined through the representation 
\begin{align}
\GL[k, \mats{k}] &=  \int_{-\infty}^{\infty} \ d \nu \ \frac{\rho_G[k,\nu]}{\mats{k}-\nu}, & \rho_G[k,\nu]&= - \frac{1}{\pi} \Im m \ \GL[k, \nu+ i 0^+]  
\llabel{spectral} 
\end{align}
where the positive definite spectral density has a representation
\beq
\rho_G[k,\nu]= \frac{1}{1- \frac{n}{2}} \ \sum_{\alpha \beta} |\langle \alpha |X^{0 \si}(k) | \beta \rangle|^2 \ (p_\alpha + p_\beta)
\delta(\nu + \varepsilon_\beta- \varepsilon_\alpha), \llabel{spectral-1}
\eeq
with $p_\alpha = \frac{1}{Z_{GC}} \exp(- \beta \varepsilon_\alpha)$, $\varepsilon_\alpha$ is an eigenvalue of the grand canonical Hamiltonian $H- \mu N$
 and $Z_{GC}$  the grand canonical partition function.
We will define the quasiparticle occupation number $m_k$ through
\begin{align} 
m_k&= \frac{1}{1- \frac{n}{2}} \langle  \ X^{\si 0 }(k) X^{0 \si}(k) \ \rangle = k_B T \sum_{\mats{k}} \ e^{ \mats{k} 0^+} \GL[k, \mats{k}] = \int d\nu \  f(\nu) \ \rho_G[k,\nu], &f(\nu)= \frac{1}{e^{\beta \nu}+1 }. \llabel{mk} 
\end{align}
Here $\si$ is  either up or down, and  $0\leq m_k \leq 1$ fixes the total number of electrons through the sum rule
\beq
\frac{1}{N_s} \sum_{\vec{k}} m_{k} = \frac{n}{2-n}. \llabel{particle-number}
\eeq
  We expect that at $T=0$, a  sharp Fermi surface of the ECQL exists generically,  and  is determined by one of several criteria in analogy with that of the Fermi liquid.  Let us list them separately now (i) The particle number sum rule Eq(\ref{particle-number}) is one of them.
(ii) A jump in $m_k$ is expected at the Fermi surface so that the locus of jumps in this defines the ``Migdal'' version of the Fermi surface. (iii)   Another is the analog of the Luttinger Ward sum rule that says that the Fermi surface  is the domain satisfying the condition $Re \GL(k,0) \geq 0$, or from the spectral representation Eq(\ref{spectral-1}) we require
\beq
 \sum_{\alpha \beta} |\langle \alpha |X^{0 \si}(k) | \beta \rangle|^2 \ \frac{p_\alpha + p_\beta}{
 \varepsilon_\beta- \varepsilon_\alpha} \geq 0. \llabel{lw-sumrule}
 \eeq
(iv) The final criterion requires that the quasiparticle life time is infinite at the FS, so that the locus of points where $\rho_{\Sigma}(k,0)=0$ defines the Fermi surface.  The approximate solutions for $\GL$ can be tested with these criteria, and we shall discuss 
only the particle number sum rule Eq(\ref{particle-number}) initially.

Below we also need the ``bubble susceptibility'' object $\chi_0[Q]$ defined through
\beq
\chi_0[Q]=\sum_q G[q]G[q+Q], \llabel{chi0}
\eeq
which can be evaluated in terms of the spectral functions readily as
\beq
\chi_0[Q]= - \frac{1}{N_s} \sum_{\vec{q}}  \int \int \ d\nu_1 \ d\nu_2 \ \rho_G[q,\nu_1] \rho_G[q+Q,\nu_2] 
\frac{f(\nu_1)-f(\nu_2)}{\nu_2-\nu_1- i \Omega_Q}.\llabel{chi0-2}
\eeq

We next describe a few systematic and consistent (i.e. Ward identity compliant) approximations that can be implemented. 
 Detailed numerical calculations within these schemes are currently under way, and the results will be presented later\cite{bss}.

\subsection{Atomic Limit}
The atomic limit is defined by switching off $t$ and $J$, and so it rather trivial. We make sure that we recover the exact answer known 
in this limit\cite{hubbard-1,hubbard-2}. We set $\Gamma_s[p_1,p_2]=\Gamma_t[p_1,p_2]=1$ and hence $\rho_G[k,\nu]= \delta(\nu+\mu)$, with $\GL^{-1}[k,\mats{k}]=
\mats{k}+\mu$, so that the chemical potential is given by
$$f(-\mu)= \frac{n}{2-n}.$$
 We compute the susceptibilities from  $\chi_s=\chi_t=\chi_0$ with
$$\chi_0[Q]= - \beta \delta_{Q,0} \ \delta_{\Omega_Q,0} \ \frac{n(1-n)}{2 (1-\frac{n}{2})^2}.$$
which agrees with the sum rule in Eq(\ref{chilocal}). 
\subsection{First approximation}
The first approximation consists of choosing $\GLO[k]$ Eq(\ref{g0-2}) and the bare vertices Eq(\ref{bare-vertex-2}). These clearly satisfy the full set of
Ward identities, In fact this approximation, with $J \to 0$ coincides with Hubbard's approximate solution, the so called  Hubbard-I solution
  of the equations of motion\cite{hubbard-1} in the limit
as $U \to \infty$. The particle number can be fixed using Eq(\ref{particle-number}), and we find that the  Fermi surface volume  encloses a fraction  $\frac{n}{2-n}$ of the first Brillouin zone, rather than the Luttinger Ward fraction of $\frac{n}{2}$. The quasiparticle number is greater than the bare particle number and this feature persists in all subsequent 
approximations. This violates the Luttinger Ward \cite{luttinger_ward} volume theorem\cite{herring},  
we discuss its implications in greater depth later in Section \ref{qp}  \footnote {The  Hubbard-I  solution\cite{hubbard-1} of the finite $U$ version
has been  criticized in literature  \cite{herring}, for failing to reproduce the Luttinger volume theorem even in  the $U \to 0$ limit.  We see 
here that while it remains incorrect for all finite values of $U$,   it 
 does  give the correct renormalized volume at  $U \to \infty$. }.

The susceptibility is just the Lindhard function $ \chi_0= - \lind[Q]$, 
\beq
 \lind[Q]= \frac{1}{N_s} \sum_{\vec{q}} \frac{ f(E^{(0)}_q-\mu)- f(E^{(0)}_{q+Q}-\mu) }{E^{(0)}_{q+Q}-E^{(0)}_q - i \Omega_Q} \llabel{lindhard}
\eeq
with $E^{(0)}_q=(1-\frac{n}{2}) \varepsilon_q$.  This function vanishes as we approach half filling $n\to 1$ and has the van Hove fingerprints of the
above defined  ``large Fermi surface''.
It is straightforward to show that this also satisfies the sum rule Eq(\ref{chi-sum-rule}), and from Eqs(\ref{physical_chi},\ref{ups_chi}),
 the physical spin susceptibility is $  2 \mu_B^2 \frac{(1-\frac{n}{2})^2}{(1-n)}\chi_L[Q]$
and the physical charge susceptibility or compressibility is $ (1-\frac{n}{2})^2 \chi_L[Q] $. Owing to the structure of the prefactors, the spin susceptibility interpolates smoothly between
the Pauli and  Curie susceptibilities on pushing the density  towards half filling $n\to 1$, while the charge susceptibility vanishes near half filling.

\subsection{  Hartree   Approximation (HA)}
The next approximation we make consists of choosing a frequency independent vertex function, and the self energy that is also frequency independent.
We write
\begin{align}
\GLI_H[k]&=(i \omega_k + \mu)- E_k^{(H)}. 
\llabel{g-hartree}
\end{align}
Specializing to only nearest neighbour hoppings, we see from symmetry 
that the form of the band dispersion remains a simple tight binding one. Therefore  for  simple cubic lattices using Eq(\ref{xt-def},\ref{afmcf},\ref{jt-js},\ref{ehat-2}) we write
\begin{align}
 E^{(H)}_k & = (1- \frac{n}{2}) \varepsilon_k  + \ehat_k =  \xi_0 \ \varepsilon_k, \nn \\
\xi_0 &\equiv \frac{1}{1-\frac{n}{2}} \left\{ (1-\frac{n}{2})^2+ \right \langle \vec{S}_{\vec{0}}\cdot \vec{S}_{\vecn} \rangle + \frac{1}{4} ( \langle n_{\vec{0}} \ n_{\vecn} \rangle-n^2)  \},\;\; \mbox{or} &\xi_0=& (1- \frac{n}{2}) + \frac{n}{4-2n} \{ 3 x_t + (1-n) x_s \}.
\llabel{e-Hartree}
\end{align}  
 The vertices are taken to be the lowest consistent ones
\begin{align}
\Gamma_s[p_1,p_2]&=1, \;\;\;  \Gamma_t[p_1,p_2]=1,   \nn \\
\Gamma_s^{\ttau}[p_1,p_2]&= (1- \frac{n}{2})(\varepsilon_{p_1}-\varepsilon_{p_2} )+ \sum_q( \varepsilon_{q+p_1}- \varepsilon_{q+p_2}) \nu[q]. 
\llabel{vertex-hartree}
\end{align}
We can check that these variables satisfy the Ward identities Eq(\ref{ward-4}) exactly.  The  object $ E_k^{(H)}$ is frequency independent, and hence to this order, the single particle spectral function is a simple delta function $
  \rho_G[k,\nu]= \delta(\nu+\mu-E^{(H)}_k)$. 
Thus $\GL^{(H)}$ has a sharp pole with no lifetime effects. The equations are non linear since the $\chi$'s are given by
\beq
\chi_{H}[Q] = \frac{1}{N_s} \sum_{\vec{Q}} \frac{f(E^{(H)}_q-\mu)-f(E^{(H)}_{q+Q}-\mu)}{E^{(H)}_{q+Q}-E^{(H)}_q- i \Omega_Q}, \llabel{hartree-chi}
\eeq
analogous to Eq(\ref{lindhard}), but with energies $E^{(H)}_k$ defined below in Eq(\ref{e-Hartree}). The Hartree energies contain a temperature dependent   renormalization of the band width, via  the  spin and charge correlation functions. 
These  nearest neighbour charge and spin correlations determine $\xi_0$, and  lead to a shrinking of the band width in case of antiferromagnetic correlations.
For ferromagnetic, i.e. Nagaoka Thouless type correlations, one has the opposite effect and magnetism promotes kinetic motion.
Near half filling the density  dependent  term in Eq(\ref{e-Hartree}) is suppressed  while the spin term survives. Due to  antiferromagnetic correlations expected at short distances, the parameter $\xi_0$ is potentially smaller than unity, temperature dependent and  can vanish, giving rise to a
metal insulator transition  The self consistent solution of $\chi_{H}[q]$  determines the spin and charge correlation functions.

\subsection{  Hartree Fock  Approximation (HFA)}
We next outline the Hartree Fock approximation where the Greens function contains the exchange term obtained from Eq(\ref{schwinger_dyson},\ref{schwinger_dyson-2}) by setting the vertices $\Gamma_s \to 1$ and $\Gamma_t \to 1$. The bare vertices are frequency independent, and all vertex corrections are functions of frequency
that vanish at high frequencies. Hence
the Hartree Fock energy is the best possible frequency independent approximation to the correct excitation energy. For this reason,  it also gives the correct form of the  first moment of the Greens function. We write
\begin{align}
\GLI_{HF}[k]&=(i \omega_k + \mu)- E_k^{(HF)}, & E_k^{(HF)} & = E_k^{(H)} - \frac{1}{2}\ \zeta_{HF} \  J_k  ,&  \zeta_{HF}=& \frac{1}{N_s} \sum_q \cos(q_x) m_q =  \langle X_{\vecn}^{\si 0} X_{0}^{0 \si} \rangle, \nn \\
\llabel{g-hartree-fock}
\end{align}
where we dropped a $k$ independent term in the self energy (from $\sum_q \varepsilon_q m_q$), and assumed the nearest neighbour hopping on a simple cubic lattice to 
simplify the expressions. The vertex functions can be written down from inspection as
\begin{eqnarray}
\Gamma_s[p_1,p_2] &=& 1 - \frac{1}{2} \sum_q J_{p_1-q} \ \chi_s[q,q+p_2-p_1] \nn \\
\Gamma_t[p_1,p_2] &=& 1 + \frac{1}{2} \sum_q J_{p_1-q} \ \chi_t[q,q+p_2-p_1] \nn \\
\Gamma_s^{\ttau}[p_1,p_2] &=& (1- \frac{n}{2})(\varepsilon_{p_1}-\varepsilon_{p_2} )+ \sum_q( \varepsilon_{q+p_1}- \varepsilon_{q+p_2}) \nu[q] - \frac{1}{2} \sum_q J_{p_1-q} \  \chi_s^{\ttau}[q,q+p_2-p_1]. \llabel{hf-vertices}
\end{eqnarray}
It is straightforward to verify that the Ward identities Eq(\ref{ward-4})  are satisfied exactly, and  so this is a consistent scheme as well. 
Unlike the earlier cases discussed, this approximation generates frequency dependent vertices.  The vertices $\Gamma_{s, \ t}[p_1,p_2]$ are 
now functions  of the momenta as well as the frequency difference $\omega_{p_2}-\omega_{p_1}$. The susceptibility
$\chi_{HF}[Q]$ can be obtained after solving the vertex functions,  the singlet and triplet susceptibilities now differ from each other. The triplet susceptibility
is enhanced at a finite value of $\vec{Q}$, whereas the singlet susceptibility is suppressed at finite $\vec{Q}$.

\subsection{ Non Linear Hartree (Fock) Approximation (NLH(F)A)  \llabel{nlha}}

The next approximation  is obtained from Eq(\ref{nlha_1}), where the terms of $O(\lambda^2)$ are isolated.
 Fourier transforming Eq(\ref{nlha_1}) we find
\normalsize
\begin{align}
 \Gamma_s[p_1,p_2] &= 1- \lambda^2 \;  \chi_s[p_2-p_1] \; \xi_s[p_1,p_2],\nn \\
 &\xi_s[p_1,p_2] = \frac{3}{16} \chi_{\loc} \{ 2 \varepsilon^t_{p_2} + J^t_{p_2-p_1} \},
 \nn \\
\Gamma_t[p_1,p_2] &= 1 +  \lambda^2 \;  \chi_t[p_2-p_1] \; \xi_t[p_1,p_2],\nn \\
 &\xi_t[p_1,p_2] = \frac{1}{16}  \;\chi_{\loc}\ \left\{  
2 \varepsilon^s_{p_2}+J^s_{p_1-p_2}  +    2 J^t_{0} - 3   J^t_{p_1-p_2} -2 \varepsilon^t_{p_1}  \right\},\nn \\
& \llabel{vertex-leading}
\end{align}
with $\chi_{loc}$ given in Eq(\ref{chilocal}). The ``$\ttau$ vertex'' follow from the stated rules, and we write
\begin{align}
 \Gamma^{\ttau}_s[p_1,p_2] &= (1- \frac{n}{2})(\varepsilon_{p_1}-\varepsilon_{p_2} )+ \sum_q( \varepsilon_{q+p_1}- \varepsilon_{q+p_2}) \nu[q] - \lambda^2 \;  \chi^{\ttau}_s[p_2-p_1] \; \xi_s[p_1,p_2]. 
 \llabel{vector-vertex-leading}
\end{align}
The partner Greens functions of the vertices Eq(\ref{vertex-leading}),  are taken from the  Hartree approximation Eq(\ref{g-hartree}).
Using  an important corollary of Eq(\ref{ward-4})
\beq
i (\omega_{p_1}- \omega_{p_2}) \chi_{s}[p_2-p_1] = \chi^{\ttau}_{s}[p_2-p_1], \llabel{ward-5}
\eeq
it is readily seen that the Ward identities Eq(\ref{ward-4}) are satisfied for these Greens functions and vertices, and thereby 
the NLHA is also a  conserving scheme.  We may add the Fock terms  to the self energy (the $\zeta$ term in Eq(\ref{g-hartree-fock}) as well as vertices (the $J$ dependent terms on the RHS of the Eq(\ref{hf-vertices})), exactly as we did in going from the Hartree to the Hartree Fock theory. This produces
the Fock generalization of the NLHA, i.e the NLHFA. We will content ourselves with a few comments about the structure of the NLHA below.

As in the case of the Hartree Fock approximation, the vertices are now frequency dependent. 
We begin by multiplying Eqs(\ref{vertex-leading}) with $\GL[p_1] \GL[p_2]$ and integrating on one of the 
two momenta to get the susceptibilities:
\beq
\chi_{s,t}[Q]= \sum_q G[q] \ \Gamma_{s,t}[q,q+Q] \ G[q+Q]. \llabel{susceptibilities}
\eeq
The answer is  given as:
\barray
(\chi_s[Q])^{-1} &=& (\chi_0[Q])^{-1} + \frac{3}{16} \lambda^2 \chi_{loc} \{ J^t_Q + 2 \ess^t_Q \} \nn \\
(\chi_t[Q])^{-1} &=& (\chi_0[Q])^{-1} - \frac{1}{16} \lambda^2 \chi_{loc} \{ 2 J^t_0- 3 J^t_Q + J^s_Q + 2( \ess^s_Q - \ess^t_Q) \}. \llabel{chimagnetic} \nn \\
\earray
using the energy type variables
\begin{align}
\ess^s_Q &=  \frac{\sum_q \varepsilon^s_{q+Q} G[q] G[q+Q] }{\chi_0[Q]}, 
 \llabel{energies-3}
\end{align}
and similarly for the triplet channel. A rough approximation is to ignore the frequency dependence of the  energy type variable $\ess^s, \ \ess^t$ and to think 
of them as further renormalized versions of the band energies and the bare exchange.  We denote these energies with bold letters, to emphasize this point, and to distinguish these from the previously defined
energies that are all real. Let us first recall from their definitions Eq(\ref{chi0},\ref{chilocal}) that both the bubble $\chi_0$ and the local susceptibility $\chi_\loc$ are negative variables. If we
treat $\ess_Q \propto \varepsilon_Q, $,  we see that the content of Eqs(\ref{chimagnetic}) is to enhance the physical  susceptibility ($  -\chi_t$) and to decrease the
physical charge compressibility ($- \chi_s$), by amounts that are sensitively dependent on the prefactors. The theory has some resemblance to the random phase approximation, but with several  coefficients including $x_s, \; x_t$  that are found self consistently.  The results will be published separately\cite{bss}.

\subsection{ Frequency dependent self energy}
From the above discussion, in all the schemes discussed so far the self energy involved in the calculations is  frequency independent. This frequency dependence is important since it provides a measure of  the decay of  quasiparticles.
 We see however that the vertices are frequency independent only  for the first few approximations, and become frequency dependent 
in the HF, NLHA and NLHFA schemes.  Since the self energy  ultimately
derives its frequency dependence from the vertex as in Eq(\ref{schwinger_dyson}, \ref{schwinger_dyson-2}), we must find a scheme where
both vertices and self energy are frequency dependent. This is an important problem to be addressed in future work. However, as a {\em  via media} solution,
we may take the frequency dependent vertices and substitute them into  Eq(\ref{schwinger_dyson-2}), as a non self consistent first approximation.
To lowest order we find terms that are reminiscent of those encountered in Fermi liquids, with $\self \propto G G G$, i.e. with  the convolution of a bubble
susceptibility with a Greens function.  This integral is a familiar one from Fermi liquid theory, and   the imaginary part of the self energy is generically  $\propto \omega^2$. The  FS volume is unchanged from the Hartree value, in a manner that is quite similar to 
the standard FL theory.   There are several detailed question that remain to be worked out regarding the shrinking of the band width and of its temperature dependence, we will return to these in future \cite{bss}

\section{Anisotropic d-wave Pairing and Superconducting instability of the ECQL \llabel{sc}}
We next study the possibility of a superconducting instability of the ECQL. We study the \tJ model without any added terms such as phonons, and hence
we are looking at the possibility of a spontaneous instability of the type that  $^3He$ undergoes, when it  becomes a superfluid\cite{ab,Leggett-1,Leggett-2}. 

Near half filling, a full Gorkov- Nambu type calculation with anomalous Greens functions within this formalism is feasible for this purpose at the level of the NLHA, and 
we will present the details later\cite{bss}. To find the 
existence of a d-wave instability, we can take a shortcut;
 following  the precedent in $^3He$,  we extract the effective interaction from the NLHA vertices Eq(\ref{vertex-leading}).
 Onsite s-wave pairing is excluded by the prohibition of double occupancy, and is automatic  in the full Gorkov Nambu type 
  scheme  for the ECQL, within our formalism\cite{bss}. However,  in the present
   phenomenological scheme, we must implement it by ignoring all but the d-wave channel\footnote{The extended s-wave channel 
   is also excluded by force in our projections, and so this treatment does not give a fair chance to that specific order.}.

The singlet and triplet vertices in Eq(\ref{vertex-leading}) are at the (NL) Hartree level, and from these we can extract the irreducible interactions\cite{nozieres}
$^1I(p,p';Q)$ and $^0I(p,p';Q)$ in the two channels, and using crossing symmetry deduce the pairing interaction in the particle particle channel.
 As Leggett points out\cite{Leggett-1}, this is achieved more transparently when we fit these interactions to a pseudopotential $U_{pp}$  treated at the same (Hartree) level\cite{nakajima}, and then consider the pairing of the pseudopotential Hamiltonian $ H= H_0+ H'_{\text{pp}}$. A strong short ranged repulsion is added 
to eliminate s-wave channel, this is necessary since we consider singlet pairing unlike $^3He$, where the triplet pairing forbids the 
s-wave channel, by symmetry. The total pseudopotential then is $ U'_{pp}(1,2)  = (\infty) \times  \delta(1,2)+ U_{pp}(1,2)  $, and the
 (spin dependent) rotationally invariant pseudopotential given by\cite{nakajima}
\barray
& & U_{pp}(1,2)  =  V(1,2) + \vec{\si}_1 \cdot \vec{\si}_2 \; W(1,2) \nn \\
&& H'_{\text{pp}} =  \frac{1}{2} \sum_{  p_1+p_2=p_3+p_4 } \langle p_1 \si_1, p_2 \si_2 | U'_{pp}|p_3 \si_3, p_4 \si_4 \rangle \;
 \hat{c}^\dagger_{p_1 \si_1} \hat{c}^\dagger_{p_2 \si_2} \hat{c}_{p_4 \si_4} \hat{c}_{p_3 \si_3} \nn \\
&&\langle p_1 \si_1, p_2 \si_2 | U_{pp}|p_3 \si_3, p_4 \si_4 \rangle = \delta_{\si_1 \si_3} \delta_{\si_2 \si_4} \{ \langle p_1 , p_2  | V |p_3 , p_4  \rangle + \si_1 \si_2 \  \langle p_1 , p_2  | W |p_3 , p_4  \rangle \} \nn \\
&& + \delta_{\si_1 \sib_3} \delta_{\si_2 \sib_4}\delta_{\si_1 \sib_2} \ \langle p_1 , p_2  | W |p_3 , p_4  \rangle, \llabel{pseudo}
\earray
where $\hat{c}$ are regarded as the Hartree quasiparticles, and we proceed to find the vertex correction  for this Hamiltonian at the Hartree level
\begin{align}
 \Gamma_s[p_1,p_2] &= 1 + 2 \sum_{p_1+p'_2=p_2+p'_1} \langle p_2 p'_1 |V | p_1 p'_2 \rangle \ \chi_s[p'_1,p'_2] \nn \\
\Gamma_t[p_1,p_2] &= 1 +  2  \sum_{p_1+p'_2=p_2+p'_1} \langle p_2 p'_1 |W | p_1 p'_2 \rangle \ \chi_t[p'_1,p'_2], 
 \llabel{vertex-pseudo}
\end{align}
where $\chi_s[p'_1,p'_2]= G[p'_1] \Gamma_s[p'_1,p'_2] G[p'_2] $ etc. Comparing with Eq(\ref{vertex-leading}) we see that
\barray
2 \langle p_2 , p'_1   | V |  p_1 , p'_2  \rangle & =& - \lambda^2 \xi_s[p_1,p_2] \nn \\
2 \langle p_2 , p'_1   | W |  p_1 , p'_2  \rangle & =& + \lambda^2 \xi_t[p_1,p_2]. 
\earray 
We can now insert these potentials into the standard anisotropic Cooper pairing problem\cite{Leggett-2,nakajima}. For the case 
of singlet pairing,  we require the pseudopotential for the process $|p_2 \uparrow - p_2 \da \rangle \to |p_1 \ua -p_1 \da \rangle  $.   We write the required matrix element
of $V-W$ 
\beq
U_{pp}(p_1,p_2) =  \frac{1}{32} \; \frac{n}{(1-n)(1-\frac{n}{2})^2} \; \left\{ 2 \ J_0 \ x_t +  ( x_s + 2 x_t) (\varepsilon_{p_1}+ \varepsilon_{p_2}) + x_s \ J_{p_1-p_2}  \right\}. \llabel{upp}
\eeq
In obtaining this expression, we have used the definitions of  $\xi_s$ and $\xi_t$ in Eq(\ref{vertex-leading}), and symmetrized these in $p_1,p_2$\cite{symmetrize}.
It is easy to see that  only the third  term  survives in the d-wave channel, and thus  we truncate further
\beq
U^{\text{d-wave}}_{pp}(p_1,p_2) =  \frac{1}{32} \; \frac{n}{(1-n)(1-\frac{n}{2})^2} \;   x_s \ J_{p_1-p_2}. \llabel{upp-2}
\eeq
Clearly the same expression holds for the full pseudopotential $U'$ since the strong repulsion has  s-wave type symmetry. 

The term $x_t J_{p_1-p_2}$ cancels out in taking $\xi_s+\xi_t$ above, and we observe that the exchange energy $J_{p_1-p_2}$ is multiplied by the nearest neighbour
density density correlations function $x_s= (\langle n_{\vecr_i} n_{\vecr_i+\vecn} \rangle - n^2)/(n -n^2)$ defined in eq(\ref{afmcf}). This object is closely 
connected to the pair distribution function discussed in the electron gas, and is well known to have a correlation hole, i.e. particles avoid getting close 
to each other regardless of their spin. In the \tJ model, a similar depletion is expected so that we expect $x_s <0$. If we consider a fully spin polarized liquid, then
we can compute $x_s$ easily  from a Fermi gas picture, and we find $x_s =- \frac{1}{n-n^2} |\langle c_0^\dagger c_{\vecn} \rangle|^2$.  This object is
negative and small near half filling $\propto - (1-n) $.
We expect $x_s$ to be negative in the ECQL in the paramagnetic limit, although the magnitude should be larger than that for the ferromagnet, since  particles need to be neighbours in order to benefit from the exchange interaction.

We now treat $U^{\text{d-wave}}_{pp}$ within the pairing scheme\cite{nakajima} and write down the gap equation for d-wave singlet superconductivity\
\barray
\Delta(p_1) & = & - \sum_{p_2} U^{\text{d-wave}}_{pp}(p_1,p_2) \ \rho(p_2) \ \Delta(p_2)\nn \\
\rho(p)&=& \left[\frac{ \tanh(\frac{1}{2} \beta E(p))}{2 E(p)}  \right] , \llabel{gap}
\earray
where $E(p)=[\xi_p^2+|\Delta(p)|^2]^{1/2}$, with the Hartree energies $\xi_p=  \xi_0 \ (\varepsilon_k - \mu_0)$,
with $\mu_0$ the chemical potential from the first approximation (without  the $\xi_0$ correction),
 and 
\beq
\xi_0= \frac{1}{1-\frac{n}{2}}\left[ (1-\frac{n}{2})^2+ \langle \vec{S}_i \cdot \vec{S}_{i + \eta} \rangle + \frac{1}{4} ( \langle n_i n_{i+\eta}\rangle-n^2)
\right]
\eeq
as defined in the NL Hartree theory Eq(\ref{e-Hartree}). This equation can be linearized near the transition temperature, by setting 
$E(p) \to |\xi_p|$ in the  summand of the above equation.  We assume  $\Delta(p) \propto  \cos(p_x)-\cos(p_y)$, and a simple analysis 
gives the condition for the transition temperature $T_c$ as 
\barray
\sum_p (\cos p_x - \cos p_y)^2 \ \frac{\tanh \left[\frac{\xi_0}{2 k_B T_c}(\varepsilon_p- \mu_0)  \right]}{2 \xi_0 (\varepsilon_p- \mu_0)} &=& \frac{1}{\alpha \ J}, 
\;\;\text{or}\nn \\
 \int_{-W}^{W} \ d\varepsilon  \rho(\varepsilon) \psi(\varepsilon) \ \frac{\tanh \left[\frac{\xi_0}{2 k_B T_c}(\varepsilon_p- \mu_0)  \right]}{2 \xi_0 (\varepsilon_p- \mu_0)} & = & \frac{1}{\alpha \ J},  \llabel{instability}
\earray
where
\barray
\psi(\varepsilon) &= &\frac{1}{\rho(\varepsilon)} \sum_p (\cos p_x - \cos p_y)^2 \delta(\varepsilon_p- \varepsilon), \;\;\text{and} \nn \\
\alpha & =& \frac{n}{ 32 (1-n)(1-\frac{n}{2})^2} \ (-x_s). \llabel{def-alpha}
\earray
 The density of states $\rho(\varepsilon)$ and the angular average $\psi(\varepsilon)$ are easily found for the square lattice in terms of the convenient variable $u=\frac{\varepsilon}{W}$ with $W= 4 t$,  and the elliptic  integrals $E(m),  K(m)$ (where $m$ is the parameter of the elliptic integrals)
\barray
\rho(\varepsilon)&=&\frac{1}{2 \pi^2 t }K(1-u^2) \nn \\
\psi(\varepsilon) &=& = 8 \left\{ \frac{1}{2}(1+u^2)- \frac{E(1-u^2)}{K(1-u^2)} \right\}.
\earray
At low temperatures, the sum diverges logarithmically from the region $\varepsilon \sim \mu_0$. We can extract the divergence by expanding
the integrand around $\mu_0$, which may be safely  taken to its   zero temperature limit. We thus find
\beq
\frac{1}{\xi_0}\rho(\mu_0) \psi(\mu_0) \log \left[1.13 \beta_c \xi_0 (W^2- \mu_0^2)^{\frac{1}{2}} \right] = \frac{1}{\alpha \ J},
\eeq
and hence 
\beq
k_B T_c \sim 1.13 \xi_0 \ \sqrt{(W^2- \mu_0^2)} \ e^{- \xi_0/\xi^*}, \llabel{tc}
\eeq
where
\beq
\xi^*= \alpha \ J \rho(\mu_0) \psi(\mu_0) .
\eeq
This expression is valid provided the resulting $Tc$ is much smaller than the band width $2 W$, and further we need the positivity of the two variables 
$\xi_0$  and $- x_s$. The maximum $T_c$ this approach can yield is $k_B T_{max} \sim 1.13 \xi^*  \sqrt{(W^2- \mu_0^2)}$. Taking standard values
for parameters in High $T_c$ systems, namely $t=6000^0$ and $J=1500^0$, this maximum $T_c$ decreases from $105^0$K at $n=.75$ to
$43^0$K at $n=.9$, provided  we use the Hartree estimates for $x_s$ and with $x_t= -.44 n $ to fit the known ground state energy of the 
Heisenberg antiferromagnet. However the $T_c$ from Eq(\ref{tc}) is much smaller than these values, 
because $\xi_0/\xi^* \gg 1$ in the entire range.

It is interesting to compare our pairing equation (\ref{gap},\ref{instability},\ref{tc}) with corresponding equations in the work of
 Baskaran, Zou and Anderson (BZA)\cite{bza}, who first proposed that the superexchange interaction could lead to superconductivity in the \tJ model,
 and to the work of Kotliar\cite{kotliar-dwave} who generalized BZA to the case of  d-wave symmetry.
  BZA's pairing equation is obtained from an intuitive argument where the exchange energy is written
 in a particular factorized  way.  Its  mean field theory results in  a pairing Hamiltonian that has a striking resemblance to our pairing term
  $U^{\text{d-wave}}_{pp}$. In fact their mean field theory transforms to  precisely to the above equations if we make the following mappings from our calculation:
$\xi_0 \to (1-n)$, $\alpha \to 1$ and finally adjust the chemical potential $\mu_0 \to \mu_{1}$, with   $\mu_{1}$  chosen
 so that $ \sum_p f(\varepsilon_p- \mu_{1})= \frac{n}{2}$. 

On the one hand the qualitative conclusions of the  two approaches are very close. Within our theory, at least within the
  NL Hartree approximation, superconductivity is possible in the d-wave channel {\em thanks to the  sign of the correlator} $x_s$; it turns  
exchange into   an attractive interaction from its initially  repulsive character. The mean field theory of BZA obtains the attraction by a
specific factorization of the exchange energy, and while it is not clear that this factorization is unique, it is consistent with the NLHA.

On the other hand, BZA attain a much greater $T_c$ than our calculation does. This can be tracked down to one slightly unfavourable and one 
crippling difference. The ratio $\xi/(1-n)$ is $O(1)$ and does not make any difference, however the ratio $\rho(\mu_{1}) \psi(\mu_{1})/\rho(\mu_0) \psi(\mu_0) $
is $O(3)$ for most of the range of densities, and this enhances their $T_c$ somewhat. The ratio $1/\alpha$ is very large $\sim O(30)$ for most densities, and this 
makes our $T_c$ come out very small. We thus see that the dimensionless constants in our expression Eq(\ref{def-alpha}) (e.g the factor of 32) make all the 
difference between the two approaches. Our approach systematically leads to these constants as stated, at least within the NLHA, and cannot be ignored.
  We thus feel that superconductivity within the ECQL is very subtle, its currently precarious  scale  could well  be influenced by correction terms beyond the NLHA considered here, and must await
further investigations.

\section{Physical Interpretation of the Quasiparticles \llabel{qp}}
We next discuss  the physical significance of the quasiparticles, defined as the poles of $\GL$,  contrasted with the bare particles obtained from $\G$. 
The number density of particles $n= N/N_s$ is found by taking the translationally invariant limit of 
\barray
 n[i]  & = & \sum_{\si} \langle \X{i}{\si \si} \rangle \ \ =  \tr \G[i^-,i] \nn \\
& = & \frac{ 1}{\gamma[i]} \{ \tr \GL[i^-,i] - 2 \det \GL[i^-,i]\} \llabel{bare-density} \\
 &=& \frac{2 \  \GL[i^-,i]}{1+\GL[i^-,i]} \;\;\;\mbox{when}\;\;\; { \W \rightarrow 0}. \llabel{number-electrons-1}
\earray
 Inverting it, we express the local quasiparticle density $n_{QP}$
\barray
n_{QP}[i] & \equiv& \tr G_{\si \si}[i^-,i],  \nn \\
 &\rightarrow & \frac{n}{1-\frac{n}{2}}, \llabel{qpdensity}
\earray
where we have taken the paramagnetic and uniform limit in the last line. Thus the QP  density is always larger than the bare density
by a factor that is  unity at very low fillings and approaches  2 near half (bare) filling.
In the case of a general spin population with the partial
 densities  denoted by $n_\si=N_\si/N_s$,   it is easy to see that 
\beq
n_{QP, \si} = \frac{n_{\si} } {1-n_{\sib} }, \llabel{number-electrons-2} 
\eeq
illustrating the fact that both spin populations of the bare electron contribute to that of the quasiparticles of a given spin. We see that the the quasiparticle densities come closer to the bare ones as we polarize the \tJ model. This is natural since  the Pauli
principle already keeps like spins apart so that the effect of the projection operators is reduced.  In the fully polarized sector the problem
 reduces to that of a spinless  ideal Fermi gas.

Fourier transforms can be performed on turning off the sources, since translation invariance is restored, and we can construct the occupation
in momentum space (see Eq(\ref{mk})) as for the standard Fermi liquid. This is carried out in detail in Eq(\ref{particle-number}).
 In the present  case, it is clear from Eq(\ref{particle-number}, \ref{number-electrons-2}) that  the number of 
  electrons ``contained'' in $\GL$ are greater than those in $\G$ by a factor $\frac{1}{1-\frac{n}{2}}$.
This is the ``renormalized'' particle number sum rule mentioned in Eq(\ref{particle-number}). From this relation we expect that the other criteria for determining the Fermi surface outlined above (after  Eq(\ref{particle-number})) are similarly scaled. Detailed calculations within various possible schemes are underway,
and we will comment  here on the basis of simple calculations. Within the Hartree or the Hartree Fock approximations, the self energy is real and the various  criteria give the same result.  The analog of the Luttinger  Ward's FS volume theorem\cite{luttinger_ward} for the extremely correlated quantum liquid (ECQL) holds, provided  we replace the electron density by an enhanced value as in Eq(\ref{number-electrons-2}).  Thus we predict Fermi surface volume $\Omega^{FS}$
for  the ECQL state, in comparison to the FL  (Fermi liquid) state and the ferromagnetic (FM)  
state (or equivalently  spinless particles) to be  given by
\begin{align}
\Omega^{FS}_{\text{ECQL}}& = \frac{n}{2-n},  &\Omega^{FS}_{\text{FL}}&=\frac{n}{2}, &\Omega^{FS}_{\text{FM}} &=n, 
&\xi^* & = \frac{\Omega^{FS}_{\text{ECQL}}}{\Omega^{FS}_{\text{FL}}}=\frac{2}{2-n}.  
 \llabel{luttinger}
\end{align}
This  renormalization of the  volume by $\xi^*$  signifies
 a lack of adiabatic continuity with the non interacting electron problem\cite{pwa-book}, a key feature of the FL. 
In Appendix \ref{hubbard-atomic}, we locate the origin of the breakdown of continuity. At least  within the limited setting of
the atomic limit of the  Hubbard model, we can trace the origin of this change in volume.  We study the change in functional dependence of the Greens function and self energy  upon cranking up the interaction
strength $U$ at a fixed frequency $\omega_n$. The distinction between two  high frequency limits: 
the weakly correlated ($\omega \to \infty \;\text{and} \; \frac{U}{\omega} \to 0 $)  or the extremely correlated  ($\omega \to \infty \;\text{and} \; \frac{U}{\omega} \to \infty $) is responsible for the changed volume of the Fermi surface.
  With the insight gleaned from this exercise,  we conjecture 
the behaviour of the general Hubbard model self energy in the limit of extreme correlations (EC). 
Assuming this conjectured behaviour,
we   provide a variation of the standard arguments\cite{luttinger_ward,agd}  that yields  the renormalized quasiparticle FS volume as in  Eq(\ref{luttinger}) 
for the Hubbard model in the \tJ regime of parameters.   These volumes satisfy the  bound
\beq 
\Omega^{FS}_{\text{FM}} \; \;  \geq \; \; \Omega^{FS}_{\text{ECQL}}\; \;   \geq\; \;  \Omega^{FS}_{\text{FL}}, \llabel{fs-bound}
\eeq
so that the ECQL  Fermi volume differs from both the standard cases for general filling, and approaches that of the FL and the FM states at low ($n\sim 0$) and high densities ($n\sim 1$) respectively.

Independently of our proposal,  unbiased numerical methods have recently   suggested that the Fermi volume of the 2-d \tJ model differs from that of the 
FL by different enlargement factors $\xi^*$ as in Eq(\ref{luttinger}), although the factors seem a bit
smaller \cite{fs_volume}.  One curious consequence follows 
for the nearest neighbour hopping bipartite lattices, e.g. the 2-d square lattice or the 3-d square lattice. At precisely $n=\frac{2}{3}$, 
the quasiparticle density  is exactly one half. Thus   the QP  FS volume is half of the first Brillouin Zone, 
and hence they occupy the  nested diamond shaped region expected for bare electrons at half filling. 
Beyond this filling, the curvature of the FS changes from  electron like to hole like.  Therefore one would expect the Hall constant of the \tJ model to change sign and become hole like at $n=\frac{2}{3}$.  Studies of the \tJ model   Hall constant\cite{sss} 
 are  consistent with this expectation,  showing a change of sign at exactly 
this filling.

It is  instructive to deduce from the quasiparticle Greens function the  time dependent number density $n_{QP}[i] = \tr G_{\si \si}[i^-,i] $ and the spin density
$\vec{S}_{QP}[i]= \frac{1}{2} \tr  \vec{\tau} \ \GL[i^-,i] $, in terms of the bare number density $n[i]$ Eq(\ref{bare-density}) and bare spin density $\vec{S}[i]= \frac{1}{2} \tr  \vec{\tau} \ \G[i^-,i] $, with  $\vec{\tau}$  the three Pauli matrices.
 These follow from the inverse relation
\beq
\GL[i^-,i]= \frac{1}{\det(\iden-\G[i^-,i])} \ (\iden-\G[i^-,i])\cdot \G[i^-,i], \llabel{g-reln} 
\eeq
and hence
\barray
n_{QP}[i] &= & 2- \frac{(1-n[i])(2-n[i])}{(1- \frac{1}{2}n[i])^2- \vec{S}[i]\cdot \vec{S}[i] } \nn \\
\vec{S}_{QP}[i]&=& \vec{S}[i] \  \frac{(1-n[i])(2-n[i])}{(1- \frac{1}{2}n[i])^2- \vec{S}[i]\cdot \vec{S}[i] }. \llabel{transform-qp}
\earray
Thus the quasi-particle number  density is locally related to the bare particle number  density and the bare spin density in a non linear fashion. 
These relations are easily inverted as well, 
\barray
n[i]&=& \frac{2 \left\{ 2+ \left(1- n_{{QP}}[i] \right)
   n_{{QP}}[i]\right\} - \sqrt{\left(2- n_{{QP}}[i]\right) \left(2- n_{{QP}}[i]- \left(2+ n_{{QP}}[i]\right) \ |\vec{S}_{QP}[i]|^2 
     \right)}}{4-(n_{{QP}}[i]{})^2 }\nn \\
\vec{S}[i]&=& \vec{S}_{QP}[i] \; \frac{1}{2-n_{QP}[i]}
\earray
 The inverse relations are in some sense more fundamental, since the quasiparticles are the basic objects
that drive  the bare particle response. We observe that  at a given number density $n[i]$, a tendency to form a local moment by the bare particles, i.e.   $|\vec{S}[i]|\rightarrow  1/2 \ n[i]$,  enhances (reduces)  the quasiparticle  spin (number) density. Further we see that the quasiparticle spin density is scaled down from the true spin density by a factor $\sim \delta = 1-n$ corresponding to the hole density measured from half filling, and also the quasiparticle band becomes full when the bare particle band becomes half filled.

It follows from Eq(\ref{transform-qp}) that the delicately structured relation ship between quasiparticle charge and spin density should be best seen when we dope the 
uniform ECQL with a charge or spin impurity. In this case the Friedel trapping   of a single bare charge ends up capturing a non trivial (density dependent) number  of quasiparticles, and the spin density reflects   the charge density as well. This leads us to expect that the role of the impurities would be important in revealing the 
nature of the quasiparticles. One interesting feature is that when the bare particle density is close to unity at any point, the quasiparticle spin density has a local minimum at that point, and thus displays a non monotonic behaviour.

From the above construction, we conclude that the charge of the quasiparticles $q_{QP}$  must be regarded as a density dependent fraction of the 
bare charge $q_e$
\beq
q_{QP}= \{1- \frac{1}{2} n\} \ q_e, \llabel{frac-charge}
\eeq
 in order that the total charge remain invariant; i.e. $q_{QP} \ N_{QP} =  q_e \ N $.  Near half filling, the charge of the quasiparticles is $\sim \frac{1}{2}$.
 Therefore the flow of a bare particle is equivalent to that  of a 
sufficient number ($\frac{2}{2-n}$)  of  quasiparticles, so that the total  charge  is balanced. These fractionally charged particles  are defined in the many body context without any specific single particle basis. These fundamentally arise in terms of   
a modified Pauli principle implied in  the  equations  
Eqs(\ref{particle-number},\ref{qpdensity},\ref{number-electrons-2}).

\section{Conclusions and Summary \llabel{sec-10}}
In this work,   we have presented a systematic study of the    \tJ  model by using the Schwinger technique of source fields. In addition we
have developed a  specific methodology to overcome the problem of non canonical Fermions forced upon us by the infinite $U$ constraint in the
model\cite{zeyher,zeyher-2}.  Since the method is technically quite involved, we have presented the details in a self contained fashion.
We obtained  the exact Schwinger Dyson equation for the \tJ model, and hence a closed expression for the inverse Greens function $\GLI[k]$ in terms  of the vertex functions.  Both singlet and triplet 
particle hole vertices are needed to
complete the definition of the $\GLI[k] $. The vertices are reported  up to the  neglect of  the source  derivatives of the vertex- this 
is a natural stopping point since we need to first understand the consequences of the many terms generated so far. In order to
facilitate concrete approximations, we presented the Ward identities for the current and density vertices.

The resulting Greens functions and vertices form a hierarchy;  this is in many respects similar to the one usually encountered for 
standard Fermi systems, but with extra features that arise from the dynamical analogs of Gutzwiller's projection operators\cite{Gutz} for the ground state. As with standard models of weakly interacting Fermions, such a hierarchy is the proper setting for exploring other features that might lead to controlled approximations. The existence of a low density nuclear matter (Brueckner, Galitskii, Migdal)  type or high density electron gas (Bohm, Pines,  Brueckner, Gell-Mann) type  approximations  in the Fermi systems are examples of such an emergent process, and are described in various texts\cite{agd,pwa-book, mahan,nozieres}. Our preliminary search shows a natural  ordering of terms in the vertex, where 
the hole density plays a central role. This scheme is currently under numerical  study and results will be reported later\cite{bss}.

The statistical mechanical equilibrium   state underlying our Greens functions is the  extremely correlated electron liquid.   This quantum liquid  breaks no spatial or temporal symmetries. It has  specific signatures that distinguish it from the Fermi liquid. In particular,
the Fermi surface volume naturally differs from that of the Fermi liquid. The elementary excitations of this liquid are best viewed as
fractionally charged quasiparticles, whose charge is determined by the density.   At all densities of the \tJ model,  the particle number sum rule 
requires a Fermi surface that is larger than the standard Luttinger-Ward Fermi surface\cite{luttinger_ward}, by an amount $\xi^*= \frac{1}{1-\frac{n}{2}}$. In Appendix \ref{hubbard-atomic}, we have presented a suggestive  variation of the standard argument for the Fermi surface volume that gives us the scaled volume.   Further studies  are needed to compute the spectral functions and thereby answer the issue of sharpness of the quasiparticles in a complete fashion, by computing $ \GLI[k]$ within a controlled scheme. 
 The fractionally charged  quasiparticle picture of the ECQL is  a quantitative
   description of the so called {\em Lower Hubbard Band}\cite{phillips}..

An interesting  feature of the theory is  that the $\GLI[k]$ involves the static density and spin correlations at nearest neighbour separation; this leads to  the narrowing of the bands and the possibility of a metal insulator transition near half filling  driven by 
local antiferromagnetic correlations (rather than true antiferromagnetic order).
 In general the ECQL is prone to various instabilities such as the antiferromagnetic, insulating or superconducting states.
We presented a calculation of the superconducting instability towards $d-wave$ order in Section \ref{sc}, the final formulae have considerable
similarity to the RVB theory\cite{bza}, but end up with a much lower $T_c$ due to some dimensionless factors that are unavoidable in this theory.
More generally, in the ECQL,  unlike  the standard Fermi liquid instabilities,  one does not need to deal with a large energy scale such as $U$,
since the Hubbard operators already deal with the local constraint efficiently. This is a great advantage since in the FL, the energy scale $U$
skews the picture of instabilities by overemphasizing the magnetic instabilities.

\section{Acknowledgments}
This work was  supported by  the grant NSF DMR 0706128.
and  the grant  DOE-BES Grant FG02-06ER46319.

\appendix

\section{Identities involving the source derivatives. \llabel{app2}}

Let us next work out the transformation of the derivative
\beq
\Delta^{-1}[i] \cdot D[j] \cdot \Delta[k] =\Delta^{-1}[i] \cdot \overline{ D[j] \cdot \Delta[k]}+ \vdots\Delta^{-1}[i] \cdot D[j] \cdot \Delta[k]\vdots \llabel{mu_def_2}
\eeq
where the first term consists of $D[j]$, the overline symbol is analogous to a contraction in field theory. In the present context it 
implies that $D$ is acting as both a matrix as well as a derivative on $\Delta[k]$. In the second term
 the vertical dots denotes ``normal ordering'' w.r.t. the derivative operator, i.e. in matrix element form, we take 
the derivative operators to {\em the  right of the $\Delta[k]$}. These two terms arise from the action of $D$ as a matrix derivative acting upon the
$\Delta[k]$ term, and also on whatever stands to the right of the expression, and is the analog of the covariant derivative in non abelian Gauge theory.
We will next show that this may be expressed as a useful identity:
\beq
\Delta^{-1}[i] \cdot D[j] \cdot \Delta[k]= \mu[i,j] \cdot \nu[j,k] + \vdots \mu[i,j] \cdot { \Delslash[j] \cdot \mu[j,k]}\vdots, \llabel{identity_3}
\eeq
where $\Delslash[j]$ is defined below in Eq(\ref{identity_1}), it is the transformed version of $D$ in terms of the transformed source term$\W$.
Let us first prove  an   identity for  the derivative. 
\barray
\vdots\Delta^{-1}[i]\cdot D[i]\cdot\Delta[i]\vdots & = & \Delslash[i] \nn \\
\Delslash_{\si_1,\si_2}[i] & = & \si_1 \si_2 \frac{\delta}{\delta \W_i^{\sib_1 \sib_2}}. \llabel{identity_1}
\earray
This useful identity implies that in calculations where the source $\U$ is turned off, as for e.g. in Appendix \ref{app9} where we calculate
the susceptibility relations, we can ignore the distinction between $D$ and $\Delslash$ since $\Delta \to \iden$.

We use Eq(\ref{delta_def}) i.e.  $\Delta^{-1}[i]=  \frac{1}{\det\Delta[i]} \Delta^k[i]$ to rewrite the inverse $\Delta$.
Taking components, we write this equation as
\barray
\left( \vdots\Delta^{-1}[i]\cdot D[i]\cdot\Delta[i]\vdots \right)_{\si_1, \si_2} & = &\frac{1}{\det\Delta[i]}\  \si_a \si_b  \Delta^k_{\si_1 \si_a}[i]
 \Delta_{\si_b \si_2}[i] \ \frac{\delta}{\delta \V_i^{\sib_a \sib_b}} \llabel{eq_temp}
\earray

Since $\W_i= \Delta^{-1}[i] \cdot \V_i \cdot \Delta[i]$,  we may write
\barray
\frac{\delta}{\delta \V_i^{\sib_a,  \sib_b}} & = & \frac{1}{\det\Delta[i]}  \Delta^k_{\sib_c \sib_a} [i] \Delta_{\sib_b \sib_d}[i] 
\frac{\delta}{\delta \W_i^{\sib_c,  \sib_d}} \nn \\
 & = & \frac{1}{\det\Delta[i]}  \Delta_{\si_a \si_c} [i] \Delta^k_{\si_d \si_b}[i] \frac{\delta}{\delta \W_i^{\sib_c,  \sib_d}}, 
\earray
where we have  using  the  definition of the conjugate in going to the second line. Substituting into Eq(\ref{eq_temp}) we see that the
spin components are now properly arranged to yield delta functions so that the identity  Eq(\ref{identity_1}) is proved.

From the identity Eq(\ref{identity_1}), we may write another useful transformation:
\beq
\vdots D[i] \vdots = \vdots\Delta[i]\cdot \Delslash[i]\cdot\Delta^{-1}[i]\vdots , \llabel{identity_2}
\eeq
this helps us to replace functional derivatives w.r.t. the original source $\V$ with those w.r.t. the transformed source $\W$.

Using this, we may at once rewrite the LHS of Eq(\ref{identity_3}) 
\barray
\vdots\Delta^{-1}[i]\cdot{ D[j]\cdot\Delta[k]}\vdots & = & \vdots\Delta^{-1}[i]\cdot\Delta[j]\cdot \Delslash[j]\cdot\Delta^{-1}[j]\cdot\Delta[k]\vdots \nn \\
&= & \vdots\mu[i,j]\cdot  \Delslash[j]\cdot\mu[j,k]\vdots 
\earray
This proves the second term in Eq(\ref{identity_3}).

\subsection{The $\nu$ matrix. \llabel{app-nu} }
We turn to the first term in Eq(\ref{identity_3}), we see from Eq(\ref{eq61})  that
\beq
\Delta[k]= \frac{1}{\gamma[k]} \left( \iden - \GL^k[k^-,k] \right),
\eeq
where we denote $\gamma[k]=1-\det \GL[k^-,k]$, therefore

\barray
\nu[k,j]& = & \Din[j]\cdot \overline{ D[j]\cdot\Delta[k]}=\Din[j]\cdot\overline{\vdots\Delta[j]\cdot\Delslash[j]\cdot\Din[j]\vdots\cdot\Delta[k] }  \nn \\
 &=&\overline{\vdots \Delslash[j]\cdot\Din[j]\vdots\cdot\Delta[k] } 
\earray
In component form we write, 
\barray
{\nu}_{\si_1,\si_2}[k,j] & = &\Din_{\si_a,\si_b}[j] \overline{ \Delslash_{\si_1,\si_a}[j] \Delta_{\si_b,\si_2}[k]}, \nn \\
&=&  \frac{\si_1 \si_a }{\gamma[k]} \Din_{\si_a,\si_b}[j] \left\{   \  \Delta_{\si_b,\si_2}[k]  \ \GL^k_{\si_q,\si_p}[k^-,k] \  
\chi_{\sib_1,\sib_a}^{\si_p,\si_q}[k,k;j] - \si_2 \si_b \; \chi_{\sib_1,\sib_a}^{\sib_2,\sib_b}[k,k;j] \right \}, \llabel{nu_def}  \  \nn \\
\earray
where we used
\beq
\frac{\delta}{\delta \W_j^{\si_a,\si_b}} \det \GL[k^-,k]=  \chi^{\si_p \si_q}_{\si_a,\si_b}[k,k;j] \  \GL^k_{\si_q,\si_p}[k^-,k].
\eeq
For spin diagonal sources, we find
\begin{align}
\nu_{\si \si}[k,j] & = \frac{1}{{\gamma [k]}}  \left[ \Din_{\si \si}[j]
   \left( \left(\Delta _{\si \si}[k]
   \GL_{\si \si}[k,k]-1 \right) \chi
   ^{(1)}[k,k,j]+\Delta
   _{\si \si }[k] \GL_{\sib \sib}[k,k] \chi
   ^{(2)}[k,k,j] \right)- \Din_{\sib \sib }[j] \chi
   ^{(3)}[k,k,j] \right].
\end{align}

Upon turning off the sources, we find
\barray
{\nu}_{\si_1,\si_2}[a,b] &=&  \delta_{\si_1,\si_2} \ \frac{1 - \frac{n}{2} }{1- n}  \{ (\frac{n}{2} -1) \chi^{(1)}[a,a;b]+ \frac{n}{2} 
\ \chi ^{(2)}[a,a;b] -   \chi ^{(3)}[a,a;b] \}. 
\earray

\subsection{The $\Theta$ matrix. \llabel{app3} }
We calculate the object
\beq
\Theta[r,s,m]= \overline{ \vdots \Delslash[r] \cdot \mu[r,s]\vdots \  \cdot \GL[s,f]}.\GLI[f,m],\llabel{theta_def_2}
\eeq
this appears above. Taking the matrix element of this we write
\barray
\Theta_{\si_1,\si_2}[r,s,m] & = & \mu_{\si_a,\si_b}[r,s] \left( \Delslash_{\si_1, \si_a}[r] \GL_{\si_b,\si_3}[s,f] \right) \GLI_{\si_3 \si_2}[f,m]
\nn \\
& = & {\si_1 \si_a} \mu_{\si_a,\si_b}[r,s] \chi_{\sib_1, \sib_a}^{\si_b,\si_3}[s,f;r]  \GLI_{\si_3 \si_2}[f,m]
\nn \\
& = & {\si_1 \si_a} \mu_{\si_a,\si_b}[r,s]  \GL_{\si_b,\si_c}[s,p] \Gamma_{\sib_1, \sib_a}^{\si_c,\si_2}[p,m;r].  \llabel{theta_def}
\earray
For spin diagonal  sources we find
\begin{align}
\Theta_{\si,\si}[r,s,m]&=\Delta _{\si \si}[s] \Din_{\si \si}[r]  \GL_{\si \si}[s,k]
   \Gamma ^{(2)}[k,m;r]-\Delta_{\sib \sib}[s] \Din_{\sib \sib}[r]
   \GL_{\sib \sib}[s,k] \Gamma^{(3)}[k,m;r]. \llabel{theta-diagonal} 
\end{align}

Upon turning off the sources, this becomes
\barray
\Theta_{\si_1,\si_2}[r,s,m]&\rightarrow& \delta_{\si_1,\si_2} G[s,k] \; \{ \Gamma
   ^{(2)}[k,m;r]-    \Gamma^{(3)}[k,m;r] \} \nn \\
&=& \delta_{\si_1,\si_2} G[s,k] \Gamma^{(p)}[k,m;r].\ \llabel{def-Theta-2}
\earray
\section{ The $\Phi_i$ matrix. \llabel{app7} }
We study the properties of the matrix $\Phi_i$ defined in Eq(\ref{eom_3}) as
\beq
\Phi_i=\Din[i] \cdot \partial_{\tau_i} \Delta[i], 
\eeq
and begin by starting from Eq(\ref{eom_anyop})
\barray
\partial_{ \tau_i} T \left( e^{- \A} \X{i}{\si_a,\si_b}(\tau_i)\right)  & = & - T \left( e^{ - \A} [  \X{i}{\si_a,\si_b}(\tau_i), H ] \right) +   \V_1^{\si_1 \si_2}( \tau_i ) 
 T \left( e^{- \A} [ \X{i}{\si_1 \si_2}(\tau_i),\X{i}{\si_a,\si_b}(\tau_i)]\right)  \;\;\mbox{ and hence}\;\nn \\
\partial_{ \tau_i} \lll \X{i}{\si_a \si_b}(\tau_i) \rrr  &=& \V_i^{\si_c \si_a}(\tau_i)   \lll \X{i}{\si_c \si_b}(\tau_i) \rrr
 -\V_i^{\si_b \si_c}(\tau_i) \lll \X{i}{\si_a \si_c}(\tau_i) \rrr
\nn \\
&& +t_{i,l} \left(  \lll \X{i}{\si_a 0}(\tau_i) \X{l}{0 \si_b}(\tau_i) \rrr - \lll \X{l}{\si_a 0}(\tau_i) \X{i}{0 \si_b}(\tau_i) \rrr \; \right) \nn \\
&& +\frac{1}{2} \; J_{i,l} \left(  \lll \X{l}{\si_a \si_c}(\tau_i) \X{i}{\si_c \si_b}(\tau_i) \rrr - \lll \X{i}{\si_a \si_c}(\tau_i) \X{l}{\si_c \si_b}(\tau_i) \rrr \; \right) 
\earray 
Thus we obtain on using  the definition $\Delta_{\si_1,\si_2}[i]= \delta_{\si_1,\si_2} - (\si_1 \si_2) \;  \lll \X{i}{\sib_1 \sib_2}(\tau_i)  \rrr$, and the $k$ conjugate of the source 
$(\V_i^{(k)})^{\si_a,\si_b}= \si_a \si_b \V_i^{\sib_b,\sib_a}$
\barray
\partial_{ \tau_i} \Delta_{\si_1,\si_2}[i] &= & ( \G^{(k)}[i,i] \cdot  \V_i^{(k)}- \V_i^{(k)}  \cdot \G^{(k)}[i,i]  ) \nn \\
&& - t_{i,l} \ (\si_1 \si_2) \    \left(  \lll \X{i}{\sib_1 0}(\tau_i) \X{l}{0 \sib_2}(\tau_i) \rrr - \lll \X{l}{\sib_1 0}(\tau_i) \X{i}{0 \sib_2}(\tau_i) \rrr \; \right) \nn \\
&& - \frac{1}{2} \; J_{i,l} \ (\si_1 \si_2) \  \left(  \lll \X{l}{\sib_1 \sib_c}(\tau_i) \X{i}{\sib_c \sib_2}(\tau_i) \rrr - \lll \X{i}{\sib_1 \sib_c}(\tau_i) \X{l}{\sib_c \sib_2}(\tau_i) \rrr \; \right) , \llabel{phi_details_1}
\earray 
and after some calculation   we find 
\barray
\Phi_i & = &  \W_i^{(k)} - \V_i^{(k)} + \bar{\Phi}_i \llabel{phi_details_3} \\
\bar{\Phi}_i & = &  t[i,l] \Din[i] \cdot  \left(\G^{(k)}[i,l] - \G^{(k)}[l,i]   \right) + \frac{1}{2} J[i,l] (\Delta[l]- \mu[i,l] \cdot \Delta[i]) \nn \\
&& + \frac{1}{2} J[i,l] \Din[i] \cdot  (\overline{ D[i] \cdot \Delta[l]} - \overline{ D[l] \cdot \Delta[i]} ). \llabel{phi_details_2}
\earray

The second and third lines of Eq(\ref{phi_details_1}) correspond to the term $\bar{\Phi}$ in Eq(\ref{phi_details_2}), and it should be noted that the
 time arguments of the $X's$ are all equal. From this it is clear that the term $\bar{\Phi}$ and vanishes on 
 turning off the sources, due to the cancellation of terms obtained by interchanging $i \leftrightarrow l$
 \footnote{  Its functional derivatives w.r.t. the sources also seems negligible at first sight, but we plan to check this more carefully  later.
 For now it must be taken as an assumption for the vertex that is certainly consistent with the Ward identity.}.

 \section{ Rotation Invariance and the Nozi\`eres relations. \llabel{app5}}
We summarize the rotational invariance argument for the various vertex functions, these are in close parallel to the standard Fermi liquid 
arguments as 
related by Nozi\`eres\footnote{This 
analysis parallels that of  Sec.6.1 for  the Bethe Salpeter equations for particle hole multiple scattering\cite{nozieres} }.  We further assume that time reversal and parity are also preserved in this putative  ECQL  state.
Let us consider  the case of an  extremely  correlated quantum liquid with no broken symmetries. Since 
 there is no source term, rotation invariance leads to a useful spin decomposition of the vertices.
A general four legged object $\chi_{\si_{c} \si_{d}}^{\si_{a}, \si_{b}}$   is analyzed initially, and 
we show that   other objects have the same decomposition. Since the symmetries of this liquid are identical to those of the 
Fermi liquid, we can extract the decomposition by studying the parallel problem. 
 We  schematically view this object as the scattering amplitude for a particle and a hole,
  from the definition of this object, it is related to the correlator (assuming a suitable set of particle times) 
\barray
\chi^{\si_{a} \si_{b}}_{\si_{c} \si_{d}}[p,q;r] & \equiv & \frac{\delta \GL_{\si_{a}\si_{b}}[p,q]}{\delta \W_r^{\si_{c} \si_{d}}}, \nn \\
& \sim& \langle f^\dagger_{\si_b}[q] f_{\si_a}[p] f^\dagger_{\si_c}[r] f_{\si_d}[r] \rangle \nn \\
& \sim& (\si_a \si_c)\langle f^\dagger_{\si_b}[q] h^\dagger_{\sib_a}[p] h_{\sib_c}[r] f_{\si_d}[r], \rangle \llabel{scattering}
\earray
where $f^\dagger_{\si}$ creates a particle with spin ${\si}$, and $h^\dagger_{\si}= f_{\sib} \ {\si} $ creates a hole   with a   spin ${\si}$. The hole creation operator has by standard convention an  extra factor of $\si$, in order to ensure that the spin flip operator of the particle $f_{\uparrow}^\dagger f_{\downarrow}$
 maps to the spin flip of the hole  $h_{\downarrow}^\dagger h_{\uparrow}$,  rather than its negative. This object is Fourier transformed according to the standard rules
Eq(\ref{fourier}), and we will omit the momentum labels for brevity.

 We may thus view the $\chi$ as a scattering amplitude for a process  taking a particle hole pair with a certain initial state to a final state\footnote{There is another way to group particles and holes leading to a second particle hole channel familiar in parquet theory, but we will choose this present one since it is most relevant.}
  as follows:
\begin{eqnarray}
\mbox{Initial state}&=& \{ \si_d, \sib_c \} \nn \\
\mbox{Final state}& =& \{ \si_b, \sib_a \}.
\end{eqnarray}
We first note that due to time reversal invariance, the object satisfies
\beq
\chi^{\si_{a} \si_{b}}_{\si_{c} \si_{d}}=\chi^{\sib_a \sib_b}_{\sib_c \sib_d},
\eeq
with reversed momenta. However, using parity, we can reverse all momenta again restoring them to their original values,
 therefore this relation is true with fixed momenta.

Let us call $\vec{S}^{tot}$ as the total spin of the particle hole pair. 
We must conserve the z component of this object 
$$S^{tot}_z= \frac{1}{2}(\sib_a+ \si_b)=\frac{1}{2}(\si_d+\sib_c).$$

We must also conserve  the total spin magnitude $|\vec{S}^{tot} | =\ 0 \mbox{ or } 1$ in this process, since 
the ground state is rotationally a singlet.
For implementing this, we further decompose the scattering into the particle hole singlet  
and triplet channels  as follows. 
 The particle hole states can be represented in one of two possible schemes
that are  illustrated for two of the states as follows:
\begin{equation}
\begin{array}{|c  c c| c  c| }\hline
&\mbox{Scheme A}&&\mbox{Scheme B}& \\ \hline
\mbox{Singlet} & \{\uparrow \downarrow - \downarrow \uparrow \}   &
 \{ f^\dagger_\uparrow  h^\dagger_\downarrow - f^\dagger_\downarrow h^\dagger_\uparrow \}  &
\{ f^\dagger_\uparrow  f_\uparrow + f^\dagger_\downarrow f_\downarrow \} &  
[ \uparrow \uparrow + \downarrow \downarrow] \\ \hline
\mbox{Triplet}& \{\uparrow \downarrow + \downarrow \uparrow \}   &
 \{ f^\dagger_\uparrow  h^\dagger_\downarrow + f^\dagger_\downarrow h^\dagger_\uparrow \}  
 &
\{ f^\dagger_\uparrow  f_\uparrow -  f^\dagger_\downarrow f_\downarrow \} 
&  [ \uparrow \uparrow - \downarrow \downarrow] \\ \hline
\end{array}
\end{equation}
Using Scheme A, we write the four possible   states (displaying the initial state) and the corresponding
scattering amplitudes  as 
\begin{equation}
\begin{array}{|c|c|c|}\hline
 &\mbox{Singlet Channel}   &  \\ 
\frac{1}{\sqrt{2}} \{ \uparrow \downarrow  + \downarrow \uparrow \} &
\chi_s   =  \frac{1}{2} \sum_{\si \si'} \chi^{\si  \si}_{\si' \si'} \rightarrow &   \chi^{(1)} + \chi^{(2)}  \\ \hline
&\mbox{Triplet Channel}  &    \\
\frac{1}{\sqrt{2}} \{ \uparrow \downarrow  - \downarrow \uparrow \}  &
 \chi_t  =   \frac{1}{2} \sum_{\si \si'} {\si \si} \chi^{\si  \si}_{\si' \si'} \rightarrow   &   \chi^{(1)} - \chi^{(2)}  \\ 
  \{ \uparrow \uparrow \}  \mbox{ or }  \{ \downarrow \downarrow \} & 
\chi_t   =    \chi^{\si  \sib}_{\si \sib}  \rightarrow & \chi^{(3)}\\ \hline
\end{array}
\end{equation}
where the    non zero amplitudes are denoted as
\begin{eqnarray}
\chi^{\si \si}_{\si \si} & \equiv & \chi^{(1)} = \frac{1}{2} \left( \chi_s+ \chi_t \right) \nn \\
\chi^{\si \si}_{\sib \sib} & \equiv & \chi^{(2)} = \frac{1}{2} \left( \chi_s- \chi_t \right)  \nn \\
\chi^{\si \sib}_{\si \sib} & \equiv & \chi^{(3)} = \chi_t.
 \end{eqnarray}

 Assuming spin isotropy of the quantum liquid phase,  and all states of a given spin
must give the same results for the scattering amplitude.
 From the two possible states  of the triplet channel, we glean an important and labour saving identity:
\beq
  \chi^{(3)}  =   \chi^{(1)} - \chi^{(2)}.  \llabel{spin_symmetry} 
\eeq
We find in the equations above that a specific combination  for the susceptibility $\chi$ and the vertex $\Gamma$ arises repeatedly:   
\barray
\chi^{(p)}&  = &   \chi_{\sib \sib}^{\sigma \sigma} -\chi_{\sib \sigma}^{\sib \sigma} =\chi^{(2)} - \chi^{(3)}, \nn \\
&=& \frac{1}{2} \{\chi_s - 3 \chi_t \}. \llabel{pairing}
\earray
This corresponds to the correlator of  {\em singlet Cooper pairs of  particles}, i.e. $\langle b^\dagger[p,q] b[r,s] \rangle $,
with $b^\dagger[p,q]= f^\dagger_{\uparrow}(p) f^\dagger_{\downarrow}(q) -  f^\dagger_{\downarrow}(p) f^\dagger_{\uparrow}(q)$, so we call it
the pair channel.
 
We also  see that the vertex function $\Gamma$ ``inherits'' this symmetry Eq(\ref{spin_symmetry}) of $\chi$.
This follows from the definition in Eq(\ref{chi_vertex})
and the isotropy and diagonal nature in spin indices of the $\GL's$ i.e. $\GL_{\si_i \si_j} \rightarrow \delta_{\si_i,\si_j}\GL_{\si_i}$ on switching off the 
source terms.

\section{ The Hubbard model atomic limit  and FS volume \llabel{hubbard-atomic}}
In this appendix, we locate the point where adiabatic continuity is lost, in going from the Hubbard model to the $t$-$J$ model. The context  of the discussion is   the atomic limit, where we can trace this quite explicitly. 
However,   the generality of the argument regarding the high frequency limit and the asymptotic form of the Greens function
and self energy
 in the two models  is to be noted.  One of the consequences is that  we have identified the 
essential  problem  in applying the Luttinger Ward theorem to the $t$-$J$ model, it must suffer corrections due to the neglect of
the boundary term $n_2$ as outlined below in Eq(\ref{errorlw}).

We recall the solution of the Hubbard model in the atomic limit\cite{hubbard-1} $t \rightarrow 0$ where
\barray
G_{\text{atomic}}(i \omega_n) & = & \frac{1- \frac{n}{2}}{i \omega_n +\mu} +  \frac{\frac{n}{2}}{i \omega_n +\mu -U}, \;\;\text{or}  \nn \\
&=& \frac{1}{i \omega_n + \mu - \Sigma(i \omega_n)}, \;\;\;\text{with} \nn \\
\Sigma(i \omega_n) & = & U \frac{n}{2} + U^2 \frac{\frac{n}{2}( 1- \frac{n}{2} )}{i \omega_n +\mu -U(1-\frac{n}{2})}. \llabel{atomic-g}
\earray
Here the symbols $G, \Sigma$ stand for the standard definitions of the Greens function and self energy\cite{agd,nozieres} and should
not be confused with the \tJ model objects defined in this work\footnote{The vertices obtained from this atomic Greens function
and the ``untreated'' self energy where the transformation Eq(\ref{gauge_transform}) has  not been done,  
are easily seen to contain linear terms at high frequency and hence are examples of a ``sick'' theory.}. 
 We are interested in 
 understanding how these functions evolve, as we go from weakly correlated (WC) to extremely correlated  (EC) limits. In the EC limit
  we recover the \tJ model solution,  
 of course in the trivial limit of $t,J \rightarrow 0$. For this we need to understand two distinct high frequency (HF) limits
\begin{align}
\text {WCHF limit}& &\omega_n& \rightarrow \infty  &U &\sim O(1) & \frac{U}{\omega_n}& \rightarrow 0 \nn \\
\text {ECHF limit} &&\omega_n   & \rightarrow \infty & U& \rightarrow \infty &\frac{U}{\omega_n}&\rightarrow \infty. \llabel{wceclimits}
\end{align}
We observe that in the WCHF limit,  $G \sim 1/i \omega_n $ and $\Sigma \sim \text{c1}+ \text{c2}/ i \omega_n $,
and thus  behave exactly as one expects\cite{agd,nozieres}. The canonical nature of the anticommutation relations of the Fermions 
fixes the coefficient of $1/i \omega_n$ in $G$
as  unity. In the EC limit, we send $U \rightarrow \infty$ first, so that 
\beq
G_{\text{atomic-EC}}( i \omega_n)=  \frac{1- \frac{n}{2}}{i \omega_n +\mu}, \llabel{atomic-g-2}
\eeq
 the coefficient of $1/i \omega_n$  is now  $1- \frac{n}{2}$ expected behaviour from  Eq(\ref{physical-greens}),
   this reflects  the non canonical nature of the electrons in this limit. This coefficient is a statement of the density dependence of the
  number of states of the ``lower Hubbard band''. The self energy in the EC limit Eq(\ref{wceclimits}) is easily found to be
 \begin{align}
 \Sigma_{EC}(i \omega_n)& = c_0 ( i \omega_n + \mu), \;\;\;\;\;\;\; \text{where}  &c_0& = - \frac{n}{2-n}. \llabel{c0}
 \end{align}
 Reflection shows that this feature of 
 a linearly growing  $\Sigma$ as a function of $i \omega_n$, an initially   surprising result, is 
 natural in this case. A linear growth of $\Sigma$  with $i \omega_n$ with the exact coefficient $c_0$ in Eq(\ref{c0}),
 is needed  to match the asymptotic high frequency behaviour of $G$ in Eq(\ref{atomic-g-2}) with the correct and related coefficient.
 Thus we see that the loss of  continuity in functional form  between weakly correlated and extremely
 correlated   Greens function $G$'s  and  $\Sigma$, arises
due of the  order of limits implied by  
 Eq(\ref{wceclimits}). The EC limit corresponds to taking $U\rightarrow \infty$ before any other limit;  this process
 tosses out a fraction of the states  of the Hilbert space, and thereby redefines the asymptotics of the remaining scales.
  We conjecture that this behaviour of the Hubbard model self energy $\Sigma$  is inevitable in the generic case of finite hopping as well.
   Thus our conjecture is that in 
  the  extreme correlations  limit of the Hubbard model,  the    self energy must have a form ($z= i \omega_n$)
  \beq
  \Sigma(\vec{k}, z)/_ {\lim_{U \rightarrow \infty}} = c_0 ( z + \mu) + \Sigma_{\text{Regular}}(\vec{k}, z), \llabel{regular}
 \eeq
with $c_0$ given in Eq(\ref{c0})
 where $\Sigma_{\text{Regular}}(\vec{k}, z)$ is a well behaved  function (i.e.$\sim c_1 + c_2/ z$ ) at high $z$. For consistency,  it must 
 in fact be related to the self energy $\self[\vec{k}, i \omega_n]$ of the \tJ model in Eq(\ref{schwinger_dyson-2}) with $J_{ij} \rightarrow 0$ through the relation
 \beq
  \Sigma_{\text{Regular}}(\vec{k}, i \omega_n) = \frac{1}{1-\frac{n}{2}}  \ \self[\vec{k}, i \omega_n]  . \llabel{relation} 
 \eeq
 
 Armed with the conjectured Eq(\ref{regular}), and another assumption detailed below, we can modify the original argument of Luttinger 
and Ward\cite{luttinger_ward} to obtain the 
 volume of the ECQL Fermi surface. Following the masters,  we begin with  the   relation between the number of particles
 and the Greens function (with  $\eta=0^+$) in a FL;
 \barray
 n & = & 2 \sum_{k} G(\vec{k}, i \omega_n) e^{i \omega_n \eta } \nn \\
 &=&  \frac{2}{N_s} \sum_{\vec{k}} \int_{- \infty}^{0} \ \frac{d x}{2 \pi i } \  \ \left \{ G(\vec{k}, x - i \eta) - G(\vec{k}, x + i \eta) \right \}.
 \earray
In the first line, the frequency sum is replaced by a contour integral after multiplying by the Fermi function. Next  we deform the contour
to run parallel to the real line, and taking the
 T=0  limit   gives the second  expression. From the definition $G(\vec{k},z)=1/(z + \mu -\varepsilon_k - \Sigma(\vec{k},z))$,
 we write 
$$G(\vec{k},z)= - \frac{d}{d z} \log G(\vec{k},z) + G(\vec{k},z) \frac{d}{d z} \Sigma(\vec{k},z),$$
so that $n= n_1+ n_2$ with
\barray
n_1&=& -   \frac{2}{N_s} \sum_{\vec{k}}  \int_{- \infty}^{0} \ \frac{d x}{2 \pi i } \  \ \left \{ \frac{d}{d x}\log G(\vec{k}, x - i \eta) - \frac{d}{d x}\log G(\vec{k}, x + i \eta) \right \} \nn \\
& = & \frac{1}{  2 \pi i}  \frac{2}{N_s} \sum_{\vec{k}} \log\frac{G(\vec{k}, -\infty - i \eta) G(\vec{k}, i \eta)}{G(\vec{k}, -\infty + i \eta) G(\vec{k}, -  i \eta)} \nn \\
&=&   \frac{2}{N_s} \sum_{\vec{k}} \Theta(G(\vec{k},0)), \llabel{aeq1}
\earray
where $\Theta(x)$  is the usual Heaviside function ($1$ or $0$). This evaluation is parallel to the one in the original argument, wherein $n_1$ is  the sole  contribution to the volume theorem. The number $n_1$,
and hence in a FL the total density $n$, is found by adding up the number per volume  of $k$ values where the $G(\vec{k},0)$ at the chemical potential is positive.
 We next consider
\barray
n_2&=&   \frac{2}{N_s} \sum_{\vec{k}}  \int_{- \infty}^{0} \ \frac{d x}{2 \pi i } \  \ \left \{ G(\vec{k}, x - i \eta) \frac{d}{d x} \Sigma(\vec{k}, x - i \eta) - G(\vec{k}, x + i \eta) \frac{d}{d x} \Sigma(\vec{k}, x + i \eta) \right \}. \llabel{errorlw} 
\earray
This term is usually integrated by parts\cite{luttinger_ward,agd} and the boundary term $\Sigma(\vec{k},x) G(\vec{k},x)$ discarded at infinity, assuming that
the growth of $\Sigma(\vec{k},x)$ is slower than linear in x. From our discussion above, {\em this assumption is incorrect for the EC or  \tJ limit}. There the linearly growing $\Sigma(\vec{k},x)$ 
precisely catches up with the inverse linearly decaying $G(\vec{k},x)$, giving a non trivial boundary contribution. The corrected answer is most easily found if we use our conjecture Eq(\ref{regular})
for decomposing $\Sigma$ into two parts, and we find
\barray
n_2& =& c_0 \  n + n_3, \;\;\;\;\text{with}  \nn \\
n_3&=&   \frac{2}{N_s} \sum_{\vec{k}} \int_{- \infty}^{0} \ \frac{d x}{2 \pi i } \  \ \left \{ G(\vec{k}, x - i \eta) \frac{d}{d x} \Sigma_{\text{Regular}}(\vec{k}, x - i \eta) - G(\vec{k}, x + i \eta) \frac{d}{d x} \Sigma_{\text{Regular}}(\vec{k}, x + i \eta) \right \}. \nn \\
\earray
Now $n_3$ can be integrated by parts safely.  We further assume with Luttinger and Ward 
that $\Sigma_{\text{Regular}}(\vec{k}, z) = \frac{\delta \Phi_{\text{Regular}}}{\delta G(\vec{k},z)}$ with a suitable  functional $\Phi_{\text{Regular}}$. 
This is the  second part of our conjecture alluded to above, and is based upon the idea that the Luttinger Ward functional $\Phi_{LW}$ itself
can be decomposed into a regular and singular parts\cite{shastry_tbp}. With this  it follows that  $n_3=0$, as shown in \cite{luttinger_ward,agd}.
  Thus we find in the  EC limit
\barray
n& = & n_1 + c_0  \ n \nn \\
 &= & (1-\frac{n}{2}) n_1,   \;\;\;\text{and thus}\nn \\
\frac{n}{1-\frac{n}{2}}&=&  \   \frac{2}{N_s} \sum_{\vec{k}} \Theta(G(\vec{k},0)), \llabel{lutt-renorm}
\earray
where we used Eq(\ref{c0}) in obtaining the second line from the first.
We emphasize that this renormalized version of the Luttinger Ward volume theorem, is only applicable in the EC limit as in Eq(\ref{wceclimits}), and
of course is the same as the FS volume for the ECQL in Eq(\ref{luttinger}).

\section{The  different limits of vanishing  source terms \llabel{app8} }
There are many possible ways to turn off the sources. The generic one used for most calculations is Case.1 below, and we also
list some other possibilities for completeness. We may distinguish three possibilities:
\begin{itemize}
\item{ Case:1}
 All source terms $$\W_i^{\si_a,\si_b} \propto \xi,$$ where $\xi$ is infinitesimal. In this case we can readily establish that
\beq
\V^{(k)}-\W^{(k)} \propto (\xi^2). 
\eeq
 The same holds for $\V- \W  \propto (\xi^2) $ and also $D-\Delslash \propto (\xi^2)$. Hence to linear order in $\xi$ we may ignore the
 distinction between $\V$ and $\W$.
 \item{ Case:2}
  Spin diagonal ${\W}$ where 
 $$\W_i^{\uparrow, \downarrow} = \W_i^{\downarrow, \uparrow }= 0. $$
In this case again we see that $\W= \V$ since all off diagonal $\G$'s vanish identically. This turns out to be not interesting since the 
$\GL$'s remain off diagonal in spin space here, because the total source term is not just $\W$ ( see next case).

\item{ Case:3} The spin polarized  case  where we choose
$$ \W_i^{\si \sib} \propto \xi ,$$
but much smaller than  $\W_i^{\uparrow  \uparrow} - \W_i^{\downarrow \downarrow }$, the latter being set to zero last.  In this case, the symmetry of the underlying ECQL state
gives us $\G_{\si,\sib} \propto \xi$, and $\G_{\si,\si}-\G_{\sib,\sib}$ is arbitrary. Thus we allow for ferromagnetic polarization in this case.
The total source term from eq(\ref{phi_details_3},\ref{eom_5}) is $\W_{tot}= W+ \W^{(k)}-  \V^{(k)}$. We compute this to first order in $\xi$ and find
\barray
(\W_{i }^{\uparrow \uparrow})_{tot} & = &\W_{i}^{\uparrow \uparrow} +\G_{\uparrow \downarrow  }[i^-,i] \frac{\W_{i}^{\uparrow  \uparrow}-\W_{i}^{\downarrow \downarrow}}{(1-\G_{\uparrow  \uparrow }[i^-,i]) (1-\G_{ \downarrow \downarrow  }[i^-,i])}  \nn \\
(\W_{i \;  }^{\downarrow  \uparrow})_{tot} & =&  \frac{(1-\G_{ \downarrow \downarrow  }[i^-,i]) \W_{i}^{ \downarrow \uparrow   } - 
\G_{ \downarrow \uparrow   }[i^-,i] (  \W_{i}^{\uparrow  \uparrow}-\W_{i}^{\downarrow \downarrow})}{1-\G_{\uparrow  \uparrow }[i^-,i] } .  \llabel{case_3} 
\earray
The other two components can be read off from the transformation  generated by reversing all spins, i.e.$\uparrow \leftrightarrow \downarrow$.
From this equation, we learn that the $\GL$'s will be diagonal in spin space if we adjust
\beq
(1-\G_{ \downarrow \downarrow  }[i^-,i]) \W_{i}^{ \downarrow \uparrow   } =
\G_{ \downarrow \uparrow   }[i^-,i] (  \W_{i}^{\uparrow  \uparrow}-\W_{i}^{\downarrow \downarrow}),
\eeq
since the $\GL$'s respond to the {\em total source} term.
\end{itemize}

\section{  Various Susceptibilities and their relationships and Sum-rules. \llabel{app9} }
We need the relationship between the response function $\ups$ of the Greens function $\G$ to the external potential $\V$ and the  function $\chi$ defined
in Eq(\ref{chi}). 

 Let us write basic identity:
\barray
\ups_{\si_c, \si_d}^{\si_a \si_b}[p,q;r] & = & \frac{\delta \G_{\si_a \si_b}[p,q]}{\delta \V_r^{\si_c \si_d}}/_{V\to 0} \nn \\
&=& (1- \frac{n}{2}) \chi_{\si_c, \si_d}^{\si_a \si_b}[p,q;r]- \si_a \si_b \ups_{\si_c, \si_d}^{\sib_b \sib_a}[p,p;r] \; \GL_{\si_b \si_b}[p,q] \llabel{ups_chi_1}
\earray
where we have used the relation $\G[p,q]= \Delta[p].\GL[p,q]$ and the relations Eq(\ref{identity_2}).
Therefore specializing to the non vanishing spin configurations we find
\barray
\ups^{(1)}[i,j;k]-(1-\frac{n}{2}) \chi^{(1)}[i,j;k]&=& -G[i,j]\; \ups^{(2)}[i^-,i;k] \nn \\
\ups^{(2)}[i,j;k]-(1-\frac{n}{2}) \chi^{(2)}[i,j;k]&=& -G[i,j]\; \ups^{(1)}[i^-,i;k] \nn \\
\ups^{(3)}[i,j;k]-(1-\frac{n}{2}) \chi^{(3)}[i,j;k]&=& +G[i,j]\; \ups^{(3)}[i^-,i;k] \nn \\
\earray 
and Fourier transforming, (with $\ups[Q]=\sum_q \ups[q,q+Q]$)
\barray
\ups^{(1)}[p_1,p_2]+ \ups^{(2)} [p_2-p_1] G[p_2] &=& (1-\frac{n}{2})\  \chi^{(1)}[p_1,p_2] \nn \\
\ups^{(2)}[p_1,p_2]+ \ups^{(1)} [p_2-p_1] G[p_2] &=& (1-\frac{n}{2})\ \chi^{(2)}[p_1,p_2] \nn \\
\ups^{(3)}[p_1,p_2]- \ups^{(3)} [p_2-p_1] G[p_2] &=& (1-\frac{n}{2})\ \chi^{(3)}[p_1,p_2] \nn \\
\earray

The susceptibilities at finite wave vectors follow by setting $p_2=p_1+Q$ and summing over $p_1$ so that
\barray
\ups^{(1)}[Q] + \frac{n}{2-n} \ups^{(2)}[Q]& = & (1-\frac{n}{2}) \chi^{(1)}[Q] \nn \\
\ups^{(2)}[Q] + \frac{n}{2-n} \ups^{(1)}[Q]& = & (1-\frac{n}{2}) \chi^{(2)}[Q] \nn \\
\ups^{(3)}[Q] - \frac{n}{2-n} \ups^{(3)}[Q]& = & (1-\frac{n}{2}) \chi^{(3)}[Q] \nn \\
\earray
so that
\begin{align}
\chi_s[Q]&= \frac{1}{(1-\frac{n}{2})^2} \ups_s[Q], &\chi_t[Q]&= \frac{1-n}{(1-\frac{n}{2})^2} \ups_t[Q]  \nn \\
 \ups_s[i^-,i;k]&= {(1-\frac{n}{2})^2}  \chi_s[i^-,i;k] , &\ups_t[i^-,i;k]&= \frac{(1-\frac{n}{2})^2}{1-n} \chi_t[i^-,i;k]  \nn \\ \llabel{ups_chi}
\end{align}

In order to gain intuition for these objects, we  note that the ``physical'' (i.e. positive definite) magnetic and charge  susceptibilities (i.e. compressibility \cite{nozieres}) at finite wave vectors are given by
\barray
\chi_{spin}[Q] & = & -  2 \mu_B^2 \ \ups_t[Q] \nn \\
\chi_{charge}[Q] & = & - \ \ups_s[Q], \nn \\ \llabel{physical_chi}
\earray
with the normalization that the corresponding objects for the (non interacting) Fermi gas are $\chi_{spin}= 2 \mu_B^2 n[0]$ and $\chi_{charge}= n[0]$ respectively, with $n[0]$ the 
density of states per spin per site at the chemical potential. Notice that the triplet object $\chi_t$ turns up with  an explicit factor $1-n$, and so it vanishes near half filling, and conversely the physical magnetic susceptibility is obtained from it by dividing with this factor.

From the definitions $\ups[i^-,i;r]= \frac{\delta \G[i^-,i]}{\delta \V_r}/_{V\to 0}$ (with suitable  spin indices),
 we can relate the susceptibilities in real space to physically interesting correlations
\barray
\ups^{(1)}[i^-,i;j] &=& - \lll \X{i}{\uparrow \uparrow}  \X{j}{\uparrow \uparrow} \rrr + \lll \X{i}{\uparrow \uparrow}\rrr \ \lll \X{j}{\uparrow \uparrow}  \rrr \nn \\
\ups^{(2)}[i^-,i;j] &=& - \lll \X{i}{\uparrow \uparrow}  \X{j}{\downarrow \downarrow} \rrr + \lll \X{i}{\uparrow \uparrow}\rrr \ \lll \X{j}{\downarrow \downarrow}  \rrr \nn \\
\ups^{(3)}[i^-,i;j] &=& - \lll \X{i}{ \uparrow \downarrow} \X{j}{ \downarrow \uparrow} \rrr + \lll \X{i}{ \uparrow \downarrow}\rrr 
\lll \X{j}{ \downarrow \uparrow} \rrr  \nn \\ 
\earray

We next  turn off the sources,  thus at equal times $\tau_i=\tau_j$, 
  we may write the correlation functions in terms of the physically meaningful charge and spin correlators:
\begin{align}
 \ups_s[i^-,i;j]&=\frac{1}{2} (n^2- \langle n_i \ n_j \rangle), &\ups_t[i^-,i;j]&= -\frac{2}{3} \langle \vec{S}_i \cdot \vec{S}_j \rangle. \llabel{ups_chi_3}
\end{align}
If we also set $i=j$, we to obtain  the  {\em local and equal time susceptibilities}
\begin{align}
\ups^{(1)}_{\loc} & = \frac{n}{2} (\frac{n}{2}-1)  & \ups^{(2)}_{\loc} & = (\frac{n}{2})^2  & \ups^{(3)}_{\loc} & = - \frac{n}{2}   \nn \\
\ups_{s,  \loc}&=  - \frac{n}{2}(1-n)  &\ups_{t, \loc}&  =  - \frac{n}{2}.& &
\end{align}
These are useful for the sum rules that we discuss next.
\subsubsection{ Susceptibility sum rules}
The local $\chi$'s follow from Eq(\ref{ups_chi}) by summing over $Q$, and these provide us with sum rules for the susceptibilities.
We find that both $\chi_s$ and $\chi_t$ satisfy {\em exactly the same sum rule}:
\begin{align}
\sum_Q \chi_s[Q]&= \frac{1}{(1-\frac{n}{2})^2} \ \sum_Q \ups_s[Q] = \chi_{\loc} \nn \\
\sum_Q \chi_t[Q]&= \frac{1-n}{(1-\frac{n}{2})^2} \ \sum_Q \ups_t[Q] = \chi_{\loc} \nn \\
\llabel{chi-sum-rule}
 \end{align}
  with
\barray
\chi_{ \loc}  &=& -\frac{n}{2}\ \frac{(1-n)}{(1-\frac{n}{2})^2}.  \llabel{chilocal}
\earray
 The vanishing with $1-n$ of both the spin and charge  $\chi$'s is an interesting  consequence of our construction.   The  equalized  local  singlet and triplet sum rules
simplify   further analysis. The two $\ups$'s are the physical susceptibilities,  relevant for   neutron scattering, NMR, charge response  and other probes.  These 
are naturally distinct from each other; for example at the insulating limit $n=1$ there is a non vanishing {\em spin response}, but no {\em charge response}.

\section{Detailed Vertices  \llabel{vertices_all}}
In this section we present the  vertices, where we have dropped the higher order vertices, i.e. set $\frac{\delta \Gamma}{\delta \W}\rightarrow 0$. 
We first break up the self energy given in Eq(\ref{schwinger_dyson_self_energy}) into convenient smaller terms, and then present the
the singlet and triplet vertex corrections arising from these six terms.
\begin{align}
 \self_1[i,j] & =     -t[i,j]  \ (\Delta[j]  -(1- \frac{n}{2})), & \self_2[i,j]&=  -t[i,j]  \nu[i,j],    \nn \\
\self_3[i,j] & =  - t[i,k]  \   \Theta[i,k,j],   &\self_4[i,j]&=   \ \delta[i,j] \frac{1}{2} J[i,k] \  \{ \mu[i,k] \cdot \Delta[i] -(1-\frac{n}{2})\}, \nn \\
\self_5[i,j] & =  + \ \delta[i,j] \ \frac{1}{2} J[i,k] \   \mu[i,k]\cdot \nu[k,i], &\self_6[i,j]&=  \frac{1}{2} J[i,k] \ \mu[i,k] \cdot \Theta[k,i,j] . \nn \\
\end{align}
We pull out the explicit factors of $\lambda= \frac{1}{1-n}$ and  
present the answers as a series in $\lambda$. The answers are given in real space with the external space time indices $i,j,m$ and four internal indices (summed over)
$a,b,c,k$.

We first  write the singlet vertices $\Gamma_s[i,j;m]_r = \frac{\delta \self_r[i,j]}{\delta V_m}$  with $1 \leq r \leq 6$:
\begin{align}
\Gamma_s[i,j;m]_1 & = \frac{1}{4} (n-2)^2 t [i,j] \chi _s [j,j,m] \nn \\
\Gamma_s[i,j;m]_2 & =\frac{3}{8} (n-2) \ \lambda \  t[i,j]
   \left((n-2) \chi _t[i,i,j] \chi
   _s[j,j,m]-2 \Gamma _t[a,b,j]
   \left(\GL[i,a] \chi
   _s[b,i,m]+\GL[b,i] \chi
   _s[i,a,m] \right)\right) \nn \\
&+\frac{1
   }{8} (n-2) t[i,j] \left(-2
   \Gamma _s[a,b,j] \left(\GL[i,a]
   \chi _s[b,i,m]+\GL[b,i] \chi
   _s[i,a,m]\right)-(n-2) \chi
   _s[i,i,j] \left(2 \chi
   _s[i,i,m]-\chi
   _s[j,j,m]\right)\right)\nn \\
&+\frac{3
   }{8} (n-2)^2 n \ \lambda^2
   t[i,j] \chi _t[i,i,j] \chi
   _s[i,i,m] \nn \\
\Gamma_s[i,j;m]_3 & =\frac{1}{4} t[i,k] \left(\Gamma
   _s[c,j,i]-3 \Gamma
   _t[c,j,i]\right) \left((n-2)
   \GL[k,c] \left(\chi
   _s[i,i,m]-\chi
   _s[k,k,m]\right)-2 \chi
   _s[k,c,m]\right) \nn \\
\Gamma_s[i,j;m]_4 & =-\frac{1}{8} (n-2)^2 \delta [i,j]
   J[i,k] \chi _s[k,k,m] \nn \\
\Gamma_s[i,j;m]_5 & =\frac{3}{8} (n-2) \ \lambda \  \delta
   [i,j] J[i,k] \left(\Gamma
   _t[a,b,i] \left(\GL[k,a] \chi
   _s[b,k,m]+\GL[b,k] \chi
   _s[k,a,m]\right)-(n-2) \chi
   _t[k,k,i] \chi
   _s[i,i,m]\right)\nn \\
&+\frac{1}{16}
   (n-2) \delta [i,j] J[i,k]
   \left(2 \Gamma _s[a,b,i]
   \left(\GL[k,a] \chi
   _s[b,k,m]+\GL[b,k] \chi
   _s[k,a,m]\right) \right) \nn \\
&-(n-2)^2 \delta [i,j] J[i,k] \left( \chi
   _s[k,k,i] \left(2 \chi
   _s[i,i,m]-3 \chi
   _s[k,k,m]\right)\right) \nn \\
& -\frac{3
   }{16} (n-2)^2 (2 n-1) \ \lambda^2 \ \delta [i,j] J[i,k] \chi
   _t[k,k,i] \chi _s[k,k,m] \nn \\
\Gamma_s[i,j;m]_6& = \frac{1}{4} J[i,k] \chi _s[i,c,m]
   \left(\Gamma _s[c,j,k]-3 \Gamma
   _t[c,j,k]\right) \llabel{app-gamma-s}
\end{align}

We next note the triplet vertices
\begin{align}
\Gamma_t[i,j;m]_1  & = -\frac{1}{4} (n-2)^2 \ \lambda \ 
   t[i,j] \chi _t[j,j,m] \nn \\
\Gamma_t[i,j;m]_2&= \left\{ \frac{1}{8} (n-2) \ \lambda \  t[i,j] \right\} \times [ \nn \\
& 2 \GL[i,a] \chi _t[b,i,m]
   \left(\Gamma _s[a,b,j]-(n+1)
   \Gamma _t[a,b,j]\right) \nn \\
& + 2 \GL[b,i] \chi _t[i,a,m]
   \left(\Gamma _s[a,b,j]-(n-3)
   \Gamma _t[a,b,j]\right) \nn \\
& -(n-2) \left(\chi _s[i,i,j] \chi
   _t[j,j,m]+2 \chi _t[i,i,j] \chi
   _t[i,i,m]\right) ] \nn \\
& -\frac{1}{8} (n-2)^2 \lambda ^2
   t[i,j] \left(n \chi _s[i,i,j]
   \chi _t[i,i,m]-\chi _t[i,i,j]
   \chi _t[j,j,m]\right) \nn \\
\Gamma_t[i,j;m]_3 &=-\frac{1}{4} (n-2) \ \lambda \  \GL[k,c]
   t[i,k] \left(\chi
   _t[i,i,m]-\chi _t[k,k,m]\right)
   \left(\Gamma _s[c,j,i]+\Gamma
   _t[c,j,i]\right) \nn \\
& -\frac{1}{2}
   t[i,k] \chi _t[k,c,m]
   \left(\Gamma _s[c,j,i]+\Gamma
   _t[c,j,i]\right) \nn \\
\Gamma_t[i,j;m]_4 &= \frac{1}{8} (n-2)^2 \ \lambda \ 
   \delta [i,j] J[i,k] \chi
   _t[k,k,m]\nn \\
\Gamma_t[i,j;m]_5 &= \left\{ \frac{1}{8} (n-2) \ \lambda \  \delta
   [i,j] J[i,k] \right\} \times [ \nn \\
& \GL[k,a] \chi _t[b,k,m] \left((n+1)
   \Gamma _t[a,b,i]-\Gamma
   _s[a,b,i]\right) \nn \\
& + (n-2) \chi _s[k,k,i] \chi
   _t[i,i,m] \nn \\
&-\GL[b,k] \chi _t[k,a,m]
   \left(\Gamma _s[a,b,i]-(n-3)
   \Gamma _t[a,b,i]\right) ]
 \nn \\
& -\frac{1}{16}
   (n-2)^2 \lambda ^2 \delta [i,j]
   J[i,k] \left(\chi _t[k,k,m]
   \left((1-2 n) \chi _s[k,k,i]+(2
   n+1) \chi _t[k,k,i]\right)-2
   \chi _t[k,k,i] \chi
   _t[i,i,m]\right) \nn \\
\Gamma_t[i,j;m]_6 &=\frac{1}{4} J[i,k] \chi _t[i,c,m]
   \left(\Gamma _s[c,j,k]+\Gamma
   _t[c,j,k]\right)-\frac{1}{2}
   (n-2) \ \lambda \  \GL[i,c] J[i,k]
   \Gamma _t[c,j,k] \left(\chi
   _t[i,i,m]-\chi _t[k,k,m]\right) \llabel{app-gamma-t}
\end{align}

\section{  Fourier transform convention. \llabel{app6} }
 Our convention for various   Fourier transforms is summarized here.
\barray
p= (\vec{p}, \omega_p), &  & r= (\vec{r}, \tau), \;\;\;\; \omega_p=  2 \pi (p+ \frac{1}{2}) k_B T \nn \\
pr &= & \vec{p}.\vec{r}- \omega_{p} \tau \nn \\
\sum_k & = & \frac{1}{N_s} \; k_B T \sum_{\omega_n}  \sum_{\vec{k}} \nn \\ 
G[a,b] &= & \sum_{k} e^{i k (a-b)} \  G[k] \nn \\
\chi[a, b;c] & = & \sum_{p_1,p_2} \chi[p_1, p_2] \; e^{i(p_1(a-c)+p_2(c-b ))} \nn \\
\chi[p_1, p_2]&=&\frac{1}{N_s}\sum_{\vecr_a,\vecr_b, \vecr_c} \; \int_0^\beta \int_0^\beta \ d(\tau_a-\tau_c)d(\tau_c-\tau_b)
  e^{- i(p_1(a-c)+p_2(c-b ))} \ \chi[a, b;c] \nn \\
\Lambda[a, b;c] & = & \sum_{p_1,p_2} \Lambda[p_1, p_2] \; e^{i(p_1(a-c)+p_2(c-b ))} \nn \\
\nu[a,b]&=& \sum_{Q} \nu(Q) \; e^{- i(Q (a-b ))} \nn \\
\chi[p_1,p_2] & = & G[p_1] \Gamma[p_1,p_2] G[p_2]  \nn \\
t[i,j] &=& - \sum_p \varepsilon_p \;\; e^{i p(r_i-r_j)} \nn \\
J[i,j] &=&  \sum_p J_p \;\; e^{i p(r_i-r_j)} \nn \\
\chi(Q)&=& \sum_p \chi(p,p+Q) \nn \\
\Upsilon[Q]& = & \sum_p \Upsilon[p,p+Q]. \llabel{fourier}
\earray


\begin{thebibliography}{99}
\bibitem{qhafm} An example of such a strong coupling liquid is the quantum Heisenberg paramagnet, at a sufficiently high  temperature. It has translation invariance and rotation invariance intact, but has no simple adiabatic relationship to a free gas with the same symmetries. The ECQL takes such a spin liquid and further introduces mobile charges into it.
\bibitem{remark_1}  In common with  standard many body theory\cite{nozieres,agd}, the Greens function is expressed in terms of a self energy, the self energy  in terms of a vertex function, and the vertex function 
 satisfies  integral equations involving  various susceptibilities and also higher order vertices.
\bibitem{remark_2} In standard perturbation theory, e.g.  as discussed in \cite{luttinger_ward,agd}  and \cite{nozieres},
 an expansion in bare Greens functions is resummed to give the  skeleton graph   expansion for the self energy $\Sigma$, or compact diagrams for the grand potential $\Omega$, where the internal lines are all the full Greens function.  The Schwinger source technique 
quite naturally generates a series that is completely analogous to this skeleton graph expansion and bypasses the first stage of bare lines. In this sense, the lack of Wick's theorem for the Hubbard operators that prevents
a regular Feynman type analysis,   is overcome by employing the Schwinger  approach.
\bibitem{symmetrize} Our choice of a left handed decomposition of $\Delta[i]$ in Eq(\ref{gauge_transform}) has the consequence that in the following theory, we will end up 
with objects, such as vertices and irreducible interactions that are not symmetric under exchanging the incoming and outgoing momenta- i.e.
we will lack manifest time reversal invariance.  One work-around is to take the resulting equations, and  to symmetrize them explicitly as need arises, as e.g. in Eq(\ref{upp}). It is also possible to work with the asymmetric objects, and unsurprisingly  lead to the same conclusions for instabilities etc. 
More directly, we can define a symmetrically factored $\G[i,f]= \sqrt{\Delta[i]} \cdot G_s[i,f] \cdot \sqrt{\Delta[f]}$, 
so that $G_s^{-1}[i,f]= \sqrt{\Delta[f]}\cdot \GLI[i,f] \cdot \sqrt{\Din[i]} $. From this it follows that the vertices $\Lambda^a$ obtained from  $G_s$ 
by differentiating w.r.t. the sources are related to 
the $\Gamma$'s used here  straightforwardly. Taking spin diagonal sources we find
via $$\Lambda^{\si \si}_{\si' \si'}[i,f;m]= \Gamma^{\si \si}_{\si' \si'}[i,f;m]- \frac{1}{2}\GLI_{\si \si}[i,f] \frac{\delta \mu_{\si \si}[i,f]}{\delta V_m^{\si' \si'}}/_{\ V \to 0}. $$
The added terms do not lead to any pathologies at high frequency.  The added terms,  explicitly available from Eq(\ref{mu-derivatives}), vanish sufficiently rapidly at high frequency to damp out the growing terms from $\GLI[i,f]$. We can then convert the asymmetric $\Gamma$ vertices to the symmetric vertices $\Lambda$ using this prescription.
\bibitem{zeyher-2} At the time of completing the mansuscript, we noticed  papers  on the \tJ model  Ref(\cite{zeyher}), with some initial overlap with our method. We differ in much of the details and the conclusions. 
 These authors consider only  density (but not  spin) dependent   sources,  and thus do not  obtain the  
 analogs  of most of our main results, e.g. the exact Schwinger Dyson equation  Eq(\ref{schwinger_dyson-2}). They further make what seem to be 
uncontrolled approximations,   leading  to results that are in disagreement with ours. For instance, they find  an effective band that vanishes as $(1-n)$ near half filling, in contrast to
our reduction of the band width by spin and density correlations.
\bibitem{shastry_tbp} We have preliminary results  showing  how  the singular part of the
self energy and the Luttinger Ward functional arise, from a partial summation of an infinite set of diagrams in perturbation theory of the 
Hubbard model.
\end{thebibliography}
\end{document}